\documentclass[10pt,reqno,a4paper]{amsart}
\usepackage[british]{babel}
\usepackage{amssymb,amstext,amsthm,eucal,bbm,mathrsfs,amscd,bbold,pifont}
\usepackage[margin=1in]{geometry}
\usepackage{array}
\usepackage[svgnames]{xcolor}
\usepackage[utf8]{inputenc}
\usepackage{enumitem}
\usepackage[small]{eulervm}
\usepackage{tgpagella}
\usepackage{tikz,pgfplots}
\usetikzlibrary{calc}
\pgfplotsset{compat=1.14}
\usepackage[unicode]{hyperref}
\hypersetup{%
  pdftitle   = {Kinematical Lie algebras via deformation theory},
  pdfkeywords = {Lie algebra, deformations, Galilean, Bargmann, central extension},
  pdfauthor  = {José Figueroa-O'Farrill},
  pdfcreator = {\LaTeX\ with package \flqq hyperref\frqq},
  linkcolor=NavyBlue,
  citecolor=ForestGreen,
  urlcolor=OrangeRed,
  anchorcolor=OrangeRed,
  colorlinks=true
}
\theoremstyle{plain}

\theoremstyle{definition}

\renewcommand{\d}{\partial}
\newcommand{\g}{\mathfrak{g}}
\newcommand{\tg}{\tilde{\mathfrak{g}}}
\newcommand{\gl}{\mathfrak{gl}}
\newcommand{\p}{\mathfrak{p}}
\newcommand{\co}{\mathfrak{co}}
\newcommand{\e}{\mathfrak{e}}

\newcommand{\so}{\mathfrak{so}}

\newcommand{\h}{\mathfrak{h}}
\renewcommand{\th}{\tilde{\mathfrak{h}}}
\newcommand{\s}{\mathfrak{s}}
\newcommand{\m}{\mathfrak{m}}
\renewcommand{\t}{\boldsymbol{t}}
\renewcommand{\u}{\boldsymbol{u}}

\newcommand{\R}{\boldsymbol{R}}
\newcommand{\B}{\boldsymbol{B}}
\newcommand{\bbeta}{\boldsymbol{\beta}}
\newcommand{\bpi}{\boldsymbol{\pi}}
\renewcommand{\P}{\boldsymbol{P}}
\newcommand{\J}{\boldsymbol{J}}
\newcommand{\K}{\boldsymbol{K}}
\newcommand{\M}{\mathcal{M}}
\renewcommand{\H}{\mathcal{H}}
\newcommand{\at}{\tilde{a}}
\newcommand{\bt}{\tilde{b}}
\newcommand{\ct}{\tilde{c}}

\newcommand{\id}{\mathbb{1}}

\let\isom=\cong
\let\tensor=\otimes
\newcommand{\RR}{\mathbb{R}}
\newcommand{\QQ}{\mathbb{Q}}

\newcommand{\PSL}{\operatorname{PSL}}
\newcommand{\Aff}{\operatorname{Aff}}
\newcommand{\GL}{\operatorname{GL}}
\newcommand{\SO}{\operatorname{SO}}
\newcommand{\Hom}{\operatorname{Hom}}
\newcommand{\tG}{\tilde{G}}
\definecolor{gris}{rgb}{0.8,0.8,0.8}
\newcommand{\zero}{{\color{gris}0}}
\allowdisplaybreaks[0]
\begin{document}

\title{Kinematical Lie algebras via deformation theory}
\author{José M. Figueroa-O'Farrill}
\address{Maxwell Institute and School of Mathematics, The University
  of Edinburgh, James Clerk Maxwell Building, Peter Guthrie Tait Road,
  Edinburgh EH9 3FD, United Kingdom}
\begin{abstract}
  We present a deformation theory approach to the classification of
  kinematical Lie algebras in $3+1$ dimensions and present
  calculations leading to the classifications of all deformations of
  the static kinematical Lie algebra and of its universal central
  extension, up to isomorphism.  In addition we determine which of
  these Lie algebras admit an invariant symmetric inner product.
  Among the new results, we find some deformations of the centrally
  extended static kinematical Lie algebra which are extensions (but
  not central) of deformations of the static kinematical Lie algebra.
  This paper lays the groundwork for two companion papers which
  present similar classifications in dimension $D + 1$ for all $D\geq
  4$ and in dimension $2+1$.
\end{abstract}
\email{j.m.figueroa@ed.ac.uk}
\thanks{EMPG-17-11}
\maketitle
\tableofcontents

\section{Introduction}
\label{sec:introduction}

The study of kinematical Lie algebras is intimately linked to the
principle of relativity, which may be interpreted as a physical avatar
of Klein's Erlanger Programme, by which a geometry can be studied via
its Lie group of automorphisms.  In the physical context, we may say
that geometrical models of the universe are determined by their
relativity group.  As in Klein's programme, by geometry one does not
necessarily mean a metric geometry, but any geometrical data which the
automorphisms leave invariant.  For example, the Newtonian model of
the universe is an affine bundle (with three-dimensional fibres to be
interpreted as \emph{space}) over an affine line (to be interpreted as
\emph{time}) and it has the galilean group as automorphisms, whose
invariant notions are the time interval between events and the
euclidean distance between simultaneous events.  By contrast,
Minkowski spacetime has the Poincaré group as the group of
automorphisms and the invariant notion is the proper distance (or,
equivalently, the proper time), which defines a lorentzian metric.
Both the galilean and Poincaré groups are examples of kinematical Lie
groups, whose Lie algebras (in dimension $3+1$) are the subject of
this paper.

Parenthetically, it is an unfortunate misnomer in Physics that we use
the word ``relativistic'' only in the case of Poincaré relativity and
``non-relativistic'' in other cases: they are all in a strict sense
relativistic, only that the relativity group is different.

By a \textbf{kinematical Lie algebra} in dimension $D$, we mean a
real $\tfrac12 (D+1)(D+2)$-dimensional Lie algebra with generators
$R_{ab} = - R_{ba}$, with $1\leq a,b \leq D$, spanning a Lie
subalgebra isomorphic to the Lie algebra $\so(D)$ of rotations in $D$
dimensions:
\begin{equation}\label{eq:so}
  [R_{ab}, R_{cd}] = \delta_{bc} R_{ad} -  \delta_{ac} R_{bd} -
  \delta_{bd} R_{ac} +  \delta_{ad} R_{bc},
\end{equation}
and $B_a$, $P_a$ and $H$ which transform according to the vector,
vector and scalar representations of $\so(D)$, respectively -- namely,
\begin{equation}
  \begin{split}
    [R_{ab}, B_c] &= \delta_{bc} B_a - \delta_{ac} B_b\\
    [R_{ab}, P_c] &= \delta_{bc} P_a - \delta_{ac} P_b\\
    [R_{ab}, H] &= 0.
  \end{split}
\end{equation}
The rest of the brackets between $B_a$, $P_a$ and $H$ are only subject
to the Jacobi identity: in particular, they must be
$\so(D)$-equivariant.  The kinematical Lie algebra where those
additional Lie brackets vanish is called the \textbf{static}
kinematical Lie algebra, of which, by definition, every other
kinematical Lie algebra is a deformation.

Up to isomorphism, there is only one kinematical Lie algebra in $D=0$:
it is one-dimensional and hence abelian.  For $D=1$, there are no
rotations and hence any three-dimensional Lie algebra is kinematical.
The classification is therefore the same as the celebrated Bianchi
classification of three-dimensional real Lie algebras \cite{Bianchi}.
The classification for $D=3$ is due to Bacry and Nuyts \cite{MR857383}
who completed earlier work of Bacry and Lévy-Leblond \cite{MR0238545}.
The present paper presents a deformation theory approach to this
classification, based on earlier work \cite{JMFGalilean} for the
galilean and Bargmann algebras, and also the classification of
deformations of the universal central extension of the static
kinematical Lie algebra.  This paper is also intended to lay the
groundwork to two further papers: in \cite{JMFKinematicalHD} we
classify the kinematical Lie algebras for $D>3$ with and without
central extension, and in \cite{TAJMFKinematical2D} we classify
kinematical Lie algebras for $D=2$.  Despite sharing the same
methodology, the problems differ sufficiently in the technicalities to
merit them being split.  A summary of the results in this series of
papers can be found in \cite{JMFKinematicalSummary}.

An important characteristic of Lie algebras, particularly for physical
applications, is whether or not they admit an invariant inner product,
by which in this paper we mean an invariant non-degenerate symmetric
bilinear form.  Such Lie algebras are said to be \textbf{metric}.
Cartan's semisimplicity criterion says that the Killing form of a
semisimple Lie algebra is an invariant inner product.  At the other
extreme, any inner product on an abelian Lie algebra is invariant.
For each of the kinematical Lie algebras in the paper we determine
which ones are metric.

This paper is organised as follows.  In
Section~\ref{sec:deform-theory-lie} we set the notation by reviewing
the basic notions of Lie algebra deformations, following in spirit the
seminal work of Nijenhuis and Richardson \cite{MR0214636}.  We recall
the definition of a graded Lie superalgebra structure on the space
$A^\bullet = \Lambda^{\bullet + 1}V^* \otimes V$ of alternating
multilinear maps from a vector space $V$ to itself and identify Lie
algebra structures on $V$ in terms of this Lie superalgebra.  We
discuss the Maurer--Cartan equation giving rise to deformations of a
Lie algebra structure on $V$ and discuss the perturbative solution of
the Maurer--Cartan equation, which lies at the heart of the
deformation-theory approach to Lie algebra classifications.  We
introduce the notions of infinitesimal deformations and of
obstructions to integrating an infinitesimal deformation and how both
can be rephrased cohomologically.  Section~\ref{sec:deform-stat-kinem}
applies this technology to recover the Bacry--Nuyts classification of
kinematical Lie algebras in dimension $3+1$.  The results are
summarised in Table~\ref{tab:kla-summary}.  A crucial role is played
by the automorphisms of the static Lie algebra which preserve the
deformation complex, so we pay considerable attention at how such
automorphisms decompose the space of cochains.  The decomposition of
the space of cochains into sub-modules of the group of automorphisms
is important in order to make a convenient choice for parametrising
the infinitesimal deformations.  The decomposition of the space of
cocycles into orbits of the group of automorphisms is crucial in the
solution of the obstruction equations and thus in determining the
integrability locus.  Finally, automorphisms play a role in bringing
deformations to normal forms so that we can determine when two
deformations are isomorphic.  In Section~\ref{sec:deform-centr-extend}
we apply this methodology to classify deformations of the universal
central extension of the static kinematical Lie algebra.  The results
here are summarised in Table~\ref{tab:defs-ce-kla-summary}: they
include some well-known Lie algebras (central extensions of
kinematical Lie algebras) and some less well-known Lie algebras which
are non-central extensions of kinematical Lie algebras.
Section~\ref{sec:conclusions} offers some conclusions.  The paper
contains three appendices.  In Appendix~\ref{sec:lie-algebra-cohom} we
review the basic notions of Chevalley--Eilenberg cohomology, whereas
in Appendices~\ref{sec:coch-stat-kinem} and
\ref{sec:coch-centr-stat-kinem} we provide details of our choice of
bases for the deformation complexes, which should allow any interested
party in reproducing our results.

\section{Deformation theory of Lie algebras}
\label{sec:deform-theory-lie}

In this section we review the basic notions of the deformation theory
of Lie algebras, first introduced by Nijenhuis and Richardson in
\cite{MR0214636}.

\subsection{The graded Lie superalgebra of alternating maps}
\label{sec:graded-lie-superalgebra}

We start by recalling the definition of a graded Lie superalgebra
structure on the space of alternating multilinear maps.

Let $V$ be a (finite-dimensional, real) vector space and let $V^*$
denote its dual.  Let $A^p = \Lambda^{p+1}V^* \otimes V$ denote the
space of skew-symmetric ($p+1$)-multilinear maps
\begin{equation}
  \underbrace{V \times \cdots \times V}_{p+1} \to V.
\end{equation}
For $\alpha \in \Lambda^{p+1}V^*$, $\beta \in \Lambda^{q+1}V^*$, and
$X,Y \in V$, let us define
\begin{equation}
  (\alpha \otimes X) \bullet (\beta \otimes Y) := (\alpha \wedge
  \iota_X \beta) \otimes Y.
\end{equation}
We then extend it bilinearly to define a product
\begin{equation}
  \bullet : A^p \times A^q \to A^{p+q}.
\end{equation}
If $\lambda \in A^p$ and $\mu \in A^q$, we define their
(Nijenhuis--Richardson) bracket by
\begin{equation}
  [\![\lambda,\mu]\!] := \lambda \bullet \mu - (-1)^{pq} \mu \bullet \lambda,
\end{equation}
which makes it clear that it is skew-symmetric (in the super-sense):
\begin{equation}
  [\![\lambda,\mu]\!] = -(-1)^{pq} [\![\mu,\lambda]\!].
\end{equation}
An easy calculation (made easier by choosing $\lambda = \alpha \otimes
X$, $\mu = \beta \otimes Y$ and $\nu = \gamma \otimes Z$), shows that
it also satisfies the Jacobi identity (also in the super-sense):
\begin{equation}
  [\![\lambda, [\![\mu,\nu]\!]]\!] = [\![[\![\lambda,\mu]\!],
  \nu]\!] + (-1)^{pq} [\![\mu, [\![\lambda,\nu]\!]]\!],
\end{equation}
where $\lambda \in A^p$ and $\mu \in A^q$.  In other words,
$(A^\bullet,[\![-,-]\!])$ is a graded Lie superalgebra.

Let us look more closely at the component $[\![-,-]\!]: A^1 \times A^1
\to A^2$.  Let $\lambda,\mu\in A^1$ and let $X,Y,Z \in V$.  Then a
short calculation shows that
\begin{equation}
  [\![\lambda,\mu]\!](X,Y,Z) =\lambda(\mu(X,Y),Z) +
  \mu(\lambda(X,Y),Z) + \text{cyclic},
\end{equation}
where, here and in the sequel, by ``cyclic'' we mean cyclic
permutations of $X,Y,Z$.  Taking $\lambda = \mu$,
\begin{equation}
  \tfrac12 [\![\mu,\mu]\!] = \mu(\mu(X,Y),Z) + \text{cyclic},
\end{equation}
whose vanishing is the Jacobi identity for the bracket on $V$ defined
by $\mu$.  In other words, $\mu \in A^1$ defines a Lie bracket on $V$
if and only if
\begin{equation}
  \label{eq:NR-Jacobi}
  [\![\mu,\mu]\!] = 0.
\end{equation}
Notice that $A^0 = V^*\otimes V$ is a Lie subalgebra under
$[\![-,-]\!]$ isomorphic to $\gl(V)$ and the adjoint
action of $A^0$ on $A^\bullet$, defined by $[\![\lambda,-]\!]$ for
$\lambda \in A^0$, is the natural action of $\gl(V)$ on $A^\bullet$.
Indeed, if $\mu \in A^p$, then
\begin{equation}
  [\![\lambda, \mu]\!](Z_0,\dots,Z_p) = \lambda \mu(Z_0,\dots,Z_p)
  - \mu(\lambda Z_0,Z_1,\dots,Z_p) - \cdots - \mu(Z_0,\dots,Z_{p-1},\lambda Z_p).
\end{equation}
This integrates to an action of $\GL(V)$, which is the group of invertible
elements in $A^0$, on $A^\bullet$ as automorphisms of the
Nijenhuis--Richardson bracket.  Therefore if $\mu$ satisfies
Equation~\eqref{eq:NR-Jacobi} so does $g\cdot\mu$, where $g \in
\GL(V)$ and
\begin{equation}
  (g\cdot \mu)(X,Y) = g \mu(g^{-1}X,g^{-1}Y).
\end{equation}
The moduli space $\M$ of Lie algebra structures on $V$ is then the
space of solutions $\mu \in A^1$ to Equation~\eqref{eq:NR-Jacobi}
modulo the action of $\GL(V)$, for it is clear that if two Lie algebra
structures are in the same orbit of $\GL(V)$ then they are isomorphic.
If a Lie algebra structure $\mu_0$ lies in the \emph{closure} of
the $\GL(V)$ orbit of a Lie algebra structure $\mu$, then $\mu$ and
$\mu_0$ may or may not give rise to isomorphic Lie algebras, but
in any case we say that the Lie algebra defined by $\mu_0$ is a
\textbf{contraction} of the Lie algebra defined by $\mu$.  To some
extent, deformation is the inverse process to contraction.

\subsection{Relationship with Chevalley--Eilenberg cohomology}
\label{sec:rel-C-E-cohom}

Let $\g$ be a Lie algebra structure on a vector space $V$ with Lie
bracket $\mu_0 \in A^1$.  Since $[\![\mu_0,\mu_0]\!]=0$, it follows from the
Jacobi identity for the Nijenhuis--Richardson bracket, that the
operation $[\![\mu_0,-]\!]: A^p \to A^{p+1}$ squares to zero:
\begin{equation}
  [\![\mu_0, [\![\mu_0, \lambda]\!]]\!] = \tfrac12
  [\![[\![\mu_0,\mu_0]\!],\lambda]\!] = 0 \qquad\text{for all~}\lambda\in A^p.
\end{equation}
In fact, it is up to a sign the Chevalley--Eilenberg differential $\d$
on the complex $C^\bullet(\g;\g)$ whose definition is recalled in
Appendix~\ref{sec:lie-algebra-cohom}.  Indeed, $[\![\mu,-]\!]$ on
$A^p$ agrees with $(-1)^p \d$ on $C^{p+1}(\g;\g)$.

Since the Chevalley--Eilenberg differential is an inner derivation
of the Nijenhuis--Richardson bracket, it follows that cocycles form a
subalgebra of the Lie superalgebra $A^\bullet$, inside which the
coboundaries form an ideal.  Therefore the Nijenhuis--Richardson
bracket descends to the cohomology and gives $H^{\bullet+1}(\g;\g)$
the structure of a graded Lie superalgebra (but with degree shifted by
one).

\subsection{Deformations of Lie algebras}
\label{sec:deform-lie-algebr}

Let us now consider deforming the Lie bracket $\mu_0$ of $\g$ to
$\mu = \mu_0 + \varphi$ for some $\varphi \in A^1$. Then $\mu$ will
define a Lie algebra if and only if Equation~\eqref{eq:NR-Jacobi} is
satisfied: \begin{equation}
  [\![\mu,\mu]\!] = [\![\mu_0 + \varphi, \mu_0 + \varphi]\!] =
  [\![\mu_0, \mu_0]\!] + 2 [\![\mu_0,\varphi]\!] +
  [\![\varphi,\varphi]\!] = 0.
\end{equation}
Since $\mu_0$ is a Lie bracket, we see that so is $\mu$ if and only if
\begin{equation}
 [\![\mu_0, \varphi]\!] + \tfrac12 [\![\varphi,\varphi]\!] = 0,
\end{equation}
and since the left-hand side is precisely $-\d\varphi$, this is equivalent to
$\varphi$ satisfying the Maurer--Cartan equation
\begin{equation}
  \d\varphi = \tfrac12 [\![\varphi,\varphi]\!].
\end{equation}

Deformation theory is essentially perturbation theory for the
Maurer--Cartan equation.  To this end we introduce a formal parameter
$t$ and write\footnote{Notice that we do not impose any convergence
  properties on the series.  We are dealing therefore with formal
  deformations.  It will turn out, however, that the deformations
  found in this paper are all polynomial and, therefore, trivially
  convergent.} $\varphi = \sum_{n\geq 1} t^n \varphi_n$, where
$\varphi_n \in A^1$ for $n=1,2,\dots$, as a formal power series in $t$
and imposing the Maurer--Cartan equation order by order in $t$.  Doing
so, we arrive at the sequence of equations:
\begin{equation}
  \label{eq:maurer-cartan-n}
  \d\varphi_n = \tfrac12   \sum_{m=1}^{n-1} [\![\varphi_m, \varphi_{n-m}]\!],
\end{equation}
for $n=1,2,\dots$.

\subsubsection{Infinitesimal deformations}
\label{sec:infin-deform}

The $n=1$ equation is simply $\d \varphi_1 = 0$, so that $\varphi_1$
is a cocycle.  Conversely, every cocycle in $C^2(\g;\g)$ defines a
\textbf{first-order} (or \textbf{infinitesimal}) deformation.  A trivial kind of infinitesimal
deformation is one which is tangent to the $\GL(V)$ orbit of $\mu_0$.  Let
$\beta \in A^0$ and consider $T = 1 + t \beta$, which is invertible
for small $t$ or as a formal power series in $t$:
\begin{equation}
  T^{-1} = 1 - t \beta + O(t^2).
\end{equation}
Then a short calculation reveals that
\begin{equation}
  (T \cdot \mu_0)(X,Y) = T \mu_0(T^{-1}X, T^{-1}Y) = (\mu_0 - t
  \d\beta)(X,Y) + O(t^2),
\end{equation}
which is an example of a deformation where $\varphi_1 = -\d\beta$ is a
coboundary. Conversely, every infinitesimal deformation which is a
coboundary is tangent to the $\GL(V)$ orbit. Therefore the
tangent space $T_{\mu_0}\M$ at $\mu_0$ to the moduli space of Lie
algebras is the quotient of infinitesimal deformations (i.e.,
cocycles) by the trivial infinitesimal deformations (i.e.,
coboundaries) and hence isomorphic to the cohomology $H^2(\g;\g)$.

\subsubsection{Obstructions to integrability}
\label{sec:obstr-integr}

Given an infinitesimal deformation $\varphi_1$, finding the
$\varphi_{n>1}$ to arrive at a deformation is known as
\textbf{integrating} $\varphi_1$.  If $\varphi_1 = -\d\beta$, then we
may just integrate it by acting with the one-parameter subgroup of
$\GL(V)$ generated by $\beta$, so integrating an infinitesimal
deformation is only ever in question when the cohomology class of
$\varphi_1$ is non-zero.

The $n=2$ equation in~\eqref{eq:maurer-cartan-n} says that
$[\![\varphi_1,\varphi_1]\!]$, which is a cocycle because $\varphi_1$
is, is actually a coboundary. In other words, the cohomology class of
$[\![\varphi_1,\varphi_1]\!]$ in $H^3(\g;\g)$ is the
\textbf{obstruction} to integrating the infinitesimal deformation
$\varphi_1$ to second order in $t$. This obstruction class only
depends on the cohomology class of $\varphi_1$ in
$H^2(\g;\g)$. Indeed, if $\varphi_1 \mapsto \varphi_1 + \d\beta$, then
\begin{equation}
  [\![\varphi_1 + \d\beta, \varphi_1 + \d\beta]\!] =
  [\![\varphi_1,\varphi_1]\!] + \d [\![\beta, \varphi_1 + \d\beta]\!].
\end{equation}

This situation persists to higher order.  Suppose that we have managed
to integrate the deformation to order $t^n$, so that
$\mu= \mu_0 + \sum_{k=1}^n t^k \varphi_k$ satisfies
\begin{equation}
  [\![\mu,\mu]\!] \in O(t^{n+1}).
\end{equation}
The Jacobi identity for the Nijenhuis--Richardson bracket says that
\begin{equation}
  [\![\mu,[\![\mu,\mu]\!]]\!]= 0,
\end{equation}
so that
\begin{equation}
 0 = [\![\mu_0 + \varphi , [\![\mu,\mu]\!]]\!] = \d[\![\mu,\mu]\!] +
 [\![\varphi, [\![\mu,\mu]\!]]\!].
\end{equation}
But $\varphi \in O(t)$ and $[\![\mu,\mu]\!] \in O(t^{n+1})$, so that
$\d[\![\mu,\mu]\!] \in O(t^{n+2})$ and hence the term of order
$t^{n+1}$ in $[\![\mu,\mu]\!]$ is a cocycle and defines a class in
$H^3(\g;\g)$.  We can integrate the deformation to order $t^{n+1}$
precisely when the class is trivial and the cocycle is a coboundary,
say, $\d\varphi_{n+1}$.

In summary, we get a sequence of obstructions in $H^3(\g;\g)$ to
integrating the infinitesimal deformation $\varphi_1$, and the
obstructions only depend on the cohomology class of $\varphi_1$ in
$H^2(\g;\g)$.

\subsubsection{Methodology}
\label{sec:methodology}

Our approach to the classification of deformations of a given Lie
algebra $\g$ will therefore consist, first of all, in calculating
$H^2(\g;\g)$.  This is simplified by the use of the Hochschild--Serre
spectral sequence, as briefly recalled in
Appendix~\ref{sec:hochschild-serre}.  For each class in $H^2(\g;\g)$
we choose a cocycle representative $\chi_i$, say, and then consider
the most general linear combination $t_1 \chi_1 + \cdots + t_N \chi_N$
of such cocycles and determine the loci in the parameter space
$\RR^N \ni (t_1,\dots,t_N)$ corresponding to the integrable
deformations.  A priori this could be an infinite process, but we will
see that all deformations in this paper are either quickly obstructed
or else integrate polynomially.  Finally, we study the action of
$\GL(V)$ on the integrable loci and pick one element from each orbit
to list the isomorphism classes of deformations.

\section{Deformations of the static kinematical Lie algebra}
\label{sec:deform-stat-kinem}

We are interested in classifying kinematical Lie algebras as
deformations of the static kinematical Lie algebra $\g$ with basis
$R_i, B_i, P_i, H $ and non-zero Lie brackets:\footnote{In $D=3$, we
  consider the rotations as vectors.  Under the dictionary is $R_{ij} =
  - \epsilon_{ijk} R_k$, the Lie bracket \eqref{eq:so} is equivalent to
  $[R_i,R_j] = \epsilon_{ijk} R_k$.}
\begin{equation}
  \label{eq:static}
  [\R, \R] = \R \qquad [\R, \B] = \B \qquad\text{and}\qquad [\R, \P] = \P
\end{equation}
in the abbreviated notation, where $[R_i,R_j] = \epsilon_{ijk} R_k$,
et cetera.  We let $\s \cong \so(3)$ denote Lie subalgebra generated
by the $R_i$ and $\h$ denote the abelian ideal spanned by
$B_i, P_i, H$.  The deformation complex $C^\bullet(\g;\g)$ is
quasi-isomorphic, by the Hochschild--Serre theorem, to the subcomplex
$C^\bullet(\h;\g)^{\s}$ of $\s$-invariant $\h$-cochains with values in
the representation $\g$.  This has the consequence that \emph{any}
deformation of the static kinematical Lie algebra is necessarily a
kinematical Lie algebra, something which was observed already in
\cite{JMFGalilean} in the context of galilean deformations.  Because
of this fact, we will work with the complex
$C^\bullet:=C^\bullet(\h;\g)^{\s}$ throughout.  Let
$\beta^i, \pi^i, \eta $ denote the basis for $\h^*$ canonically dual
to $B_i,P_i,H$, respectively.  The Chevalley--Eilenberg differential
$\d : C^p \to C^{p+1}$ defines subspaces $Z^p \subset C^p$ of cocycles
and $B^p \subset C^p$ of coboundaries.  The cohomology $H^p = Z^p/B^p$
is not a subspace of $C^p$, but we may identify it with a subspace of
$C^p$ by making a choice of cocycle representative for each element in
a basis of $H^p$.  Let $\H^p \subset C^p$ be such a choice, so that we
may decompose $Z^p = B^p \oplus \H^p$.  A convenient choice is one
where $\H^p$ is stable under those automorphisms of $\h$ which commute
with the action of $\s$.

\subsection{Automorphisms of $\h$}
\label{sec:automorphisms-h}

Since $\h$ is abelian, the automorphism group is the general linear
group $\GL(\h)$.  However we are only interested in those
automorphisms which commute with the action of $\s$, so that they act
on the $\s$-invariant deformation complex.  To this end we will
consider the group $G = \RR^\times \times \GL(\RR^2)$ acting on $\h$ in
the following way.  If $A = \begin{pmatrix}a & b \\ c & d
\end{pmatrix} \in \GL(\RR^2)$ and $\lambda \in \RR^\times$, then
\begin{equation}
  (\B,\P,H ) \mapsto (\B, \P, H )
  \begin{pmatrix} 
    a & b & \zero \\ c & d & \zero \\ \zero & \zero & \lambda
  \end{pmatrix} = (a \B + c \P, b \B + d \P, \lambda H ).
\end{equation}
The induced action on $\h^*$ is
\begin{equation}
  \bbeta \mapsto \Delta^{-1}(d \bbeta - b \bpi) \qquad \bpi \mapsto
  \Delta^{-1}(-c \bbeta + a \bpi) \qquad\text{and}\qquad \eta  \mapsto \lambda^{-1} \eta ,
\end{equation}
where $\Delta = \det A = ad - bc $.

\subsection{Infinitesimal deformations}
\label{sec:inf-def}

In the notation of Appendix~\ref{sec:coch-stat-kinem} and in
particular from the action of the Chevalley--Eilenberg differential
given by Equation~\eqref{eq:CE-static}, we see that the spaces of
coboundaries $B^2$ and cocycles $Z^2$ are given by
\begin{equation}
  B^2 = \RR\left<2 c_9 + c_{13}, c_{12} + 2 c_{16} \right>
  \qquad\text{and}\qquad Z^2 =
  \RR\left<c_1,c_3,c_4,c_6,c_7,c_9,c_{10},c_{12},c_{13},c_{15},c_{16}\right>,
\end{equation}
where the notation $\RR\left<...\right>$ means the real subspace
spanned by the vectors inside the angle brackets.  Under the
$G$-action, $Z^2$ decomposes into the following submodules:
\begin{equation}
  Z^2 = \RR\left<c_1\right> \oplus \RR\left<c_3 + c_7\right> \oplus
  \RR\left<c_3 - c_7, c_4, c_6\right> \oplus
  \RR\left<c_9-c_{13},c_{10},c_{12}-c_{16},c_{15}\right> \oplus B^2,
\end{equation}
so that the cohomology $H^2$ is isomorphic (as a $G$-module) to the
subspace $\H^2 \subset Z^2$ defined by
\begin{equation}
  \H^2 = \RR\left<c_1\right> \oplus \RR\left<c_3 + c_7\right> \oplus
  \RR\left<c_3 - c_7, c_4, c_6\right> \oplus
  \RR\left<c_9-c_{13},c_{10},c_{12}-c_{16},c_{15}\right>.
\end{equation}
Therefore the most general (non-trivial) infinitesimal deformation can
be parametrised as
\begin{equation}
  \label{eq:inf-def}
  \varphi_1 = t_1 c_1 + t_2 (c_3 + c_7) + t_3 c_4 + t_4 c_6 + t_5 (c_3
  - c_7) + t_6 c_{10} + t_7 (c_{12}-c_{16}) + t_8 (c_9-c_{13}) + t_9
  c_{15}.
\end{equation}

\subsection{Obstructions}
\label{sec:obstructions}

The first obstruction is the class of $\tfrac12
[\![\varphi_1,\varphi_1]\!]$ in $H^3$.  Using the notation in
Appendix~\ref{sec:coch-stat-kinem} and in particular the determination
of the Nijenhuis--Richardson bracket \eqref{eq:NR-static-explicit}, we
calculate
\begin{equation}
  \begin{split}
    \tfrac12 [\![\varphi_1,\varphi_1]\!] &= b_1 (t_1 t_3 + 2 t_6 t_7 +
    2 t_8^2) + b_2 (-\tfrac12 t_4 t_6 + t_3 t_7 + \tfrac32 t_2 t_8 +
    \tfrac32 t_5 t_8)\\
    & \quad {} + b_3 (t_4 t_6 - 2 t_3 t_7 - t_2 t_8 - t_5 t_8) + b_4
    (t_2 t_7 - t_5 t_7 + 2 t_4 t_8 + t_3 t_9)\\
    & \quad {} + 
    \tfrac12 b_5 (- t_2 t_7 + t_5 t_7 - 2 t_4 t_8 - t_3 t_9) +
    \tfrac12 b_6 (t_2 t_6 + 3 t_5 t_6 - t_3 t_8)\\
    & \quad {} + b_8 (t_1 t_2 - t_1 t_5  + t_7 t_8 + t_6 t_9) + b_9
    (t_1 t_2 + t_1 t_5 - t_7 t_8 - t_6 t_9)\\
    & \quad {} + b_{11} (t_1 t_4 - 2 t_7^2 + 2 t_8 t_9) +
    b_{12}(-\tfrac12 t_1 t_6) + 2 b_{13} t_1 t_2 + \tfrac12 b_{14} (3
    t_4 t_7 + t_2 t_9 - 3 t_5 t_9)\\
    & \quad {} + b_{18} (-\tfrac12 t_1 t_8) + b_{19} (t_1 t_7) +
    b_{20} (\tfrac12 t_1 t_7 + t_1 t_9).
  \end{split}
\end{equation}
We know that this is a cocycle in $Z^3$ and from
Equation~\eqref{eq:CE-static} we know that $B^3$ is spanned by $b_1,
b_2 + b_3, b_4 + b_5, b_8 - b_9, b_{11}$.  In other words, if we let
$[-]$ denote the class in $H^3$ of a cocycle, we see that
\begin{equation}
  \begin{split}
    \left[ \tfrac12 [\![\varphi_1,\varphi_1]\!]  \right] &= [b_2] (
    -\tfrac32 t_4 t_6 + 3 t_3 t_7 + \tfrac52 t_2 t_8 + \tfrac52 t_5 t_8) + [b_4] (
    \tfrac32 t_2 t_7 - \tfrac32 t_5 t_7 + 3 t_4 t_8 + \tfrac32 t_3 t_9) \\
    & \quad {} + \tfrac12 [b_6] (t_2 t_6 + 3 t_5 t_6 - t_3 t_8)
    + 2 [b_8] t_1 t_2 - \tfrac12 [b_{12}] t_1 t_6 +
    2 [b_{13}] t_1 t_2 \\
    &\quad {} + \tfrac12 [b_{14}] (3 t_4 t_7 + t_2 t_9 - 3 t_5 t_9) -
    \tfrac12 [b_{18}] t_1 t_8 + [b_{19}] t_1 t_7 +
    [b_{20}] (\tfrac12 t_1 t_7 + t_1 t_9).
  \end{split}
\end{equation}
This obstruction class is zero on the intersection of the following 9
quadrics\footnote{A Gröbner basis for the ideal of
  $\QQ[t_1,\dots,t_9]$ generated by this system of quadrics has 17
  polynomials of degrees ranging from 2 to 5.  Although it is possible
  to solve these polynomial equations and find all branches of their
  zero locus, we prefer a less black-boxy approach.}
\begin{equation}
  \label{eq:quad-obstr}
  \begin{aligned}[m]
    t_1 t_2 &= 0\\
    t_1 t_6 &= 0\\
    t_1 t_7 &= 0\\
    t_1 t_8 &= 0\\
    t_1 t_9 &= 0
  \end{aligned}
  \qquad\qquad
  \begin{aligned}[m]
    -3 t_4 t_6 + 6 t_3 t_7 + 5 t_2 t_8 + 5 t_5 t_8 &= 0\\
    t_2 t_7 - t_5 t_7 + 2 t_4 t_8 + t_3 t_9 &= 0\\
    t_2 t_6 + 3 t_5 t_6 - t_3 t_8 &= 0\\
    3 t_4 t_7 + t_2 t_9 - 3 t_5 t_9 &=0.\\
  \end{aligned}
\end{equation}
Provided that these quadratic equations are satisfied, we find that
$\tfrac12[\![\varphi_1,\varphi_1]\!] = \d\varphi_2$, where
\begin{equation}
    \varphi_2 = \tfrac25 ( 2 t_3 t_7 - t_4 t_6 ) c_2 + (t_1 t_3 + 2
    t_6 t_7 + 2 t_8^2) c_8 + (t_1 t_2 - t_1 t_5 + t_7 t_8 + t_6 t_9) c_{11} -
    (t_1 t_4 - 2 t_7^2 + 2 t_8 t_9) c_{14}.
\end{equation}

The next obstruction is $[\![\varphi_1,\varphi_2]\!]$.  Provided that
the quadratic obstructions \eqref{eq:quad-obstr} are satisfied, it is
given by
\begin{multline}
  [\![\varphi_1,\varphi_2]\!] = b_7 (-t_4 t_6 t_8 ) + \tfrac12 b_{10}
  (6 t_4 t_6 t_7 + 5 t_2 t_7 t_8 - 5 t_5 t_7 t_8 + 10 t_4 t_8^2 + 3
  t_2 t_6 t_9 - 3 t_5 t_6 t_9)\\ +
  \tfrac15 b_{17} (12 t_2 t_7^2 - 12 t_5 t_7^2 + 9 t_4 t_7 t_8 + 6 t_4 t_6 t_9 -10
  t_2 t_8 t_9 + 10 t_5 t_8 t_9).
\end{multline}
The class in $H^3$ corresponding to this cocycle vanishes if and only
if each of the coefficients of $b_7$, $b_{10}$, and $b_{17}$ vanish,
yielding the following system of cubics:
\begin{equation}
  \label{eq:cubic-obstr}
  \begin{split}
    t_4 t_6 t_8 &= 0\\
    2 t_4 (3 t_6 t_7 + 5 t_8^2) + (t_2 - t_5) (5 t_7 t_8 +  3 t_6 t_9) &=  0\\
    3 t_4 (3 t_7 t_8 + 2 t_6 t_9) + 2 (t_2 - t_5) (6 t_7^2 - 5 t_8 t_9) &= 0.
  \end{split} 
\end{equation}
If this is the case, $[\![\varphi_1,\varphi_2]\!]$ vanishes (on the
nose and not just in cohomology), so we can take $\varphi_3 = 0$.  The
next obstruction is $\tfrac12 [\![\varphi_2,\varphi_2]\!]$, which also
vanishes, since from \eqref{eq:NR-static-explicit} we see that
$c_2, c_8, c_{11}, c_{14}$ are contained in an abelian subalgebra of
the Nijenhuis--Richardson superalgebra.  Therefore we can take
$\varphi_4 = 0$ as well.  There are thus no further obstructions and
therefore the infinitesimal deformation is integrable on the combined
locus of the system~\eqref{eq:quad-obstr} of quadrics and the
system~\eqref{eq:cubic-obstr} of cubics.  We could hit this system
with the Gröbner hammer, but we prefer to exploit the action of the
automorphisms in order to solve it in a more transparent fashion.

\subsection{The action of automorphisms on the deformation parameters}
\label{sec:acti-autom-deform}

The group $G = \RR^\times \times \GL(2,\RR)$ acts on the
nine-dimensional vector space $\H^2$ and in particular this induces an
action on the coordinates $t_1,\dots,t_9$ which parametrise it.
Rather than tacking the action of $G$ on this nine-dimensional space,
it is computationally convenient to focus on how $G$ acts in a
subspace of smaller dimension whose orbit structure is easier to
determine.  To this end, let us focus on the three-dimensional
subspace spanned by $t_3, t_4, t_5$.  Then the action of
$(\lambda, A) \in G$ on $ (t_3, t_4, t_5)$ is
\begin{equation}
  \begin{pmatrix}
    t_3 \\ t_4  \\ t_5
  \end{pmatrix}
  \mapsto \frac1{\lambda \Delta}
  \begin{pmatrix}
    d^2 & -c^2 & 2cd \\ -b^2 & a^2 & -2 a b \\ b d & -a c & ad + bc
  \end{pmatrix}
  \begin{pmatrix}
    t_3 \\ t_4  \\ t_5
  \end{pmatrix}= \frac1\lambda M_A   \begin{pmatrix}
    t_3 \\ t_4  \\ t_5
  \end{pmatrix},
\end{equation}
which defines $M_A$ and where
$A = \begin{pmatrix} a & b \\ c & d \end{pmatrix}$ and
$\Delta = \det A$.  The kernel of this three-dimensional
representation of $G$ consists of
$\left\{(1,\mu \id)~ \middle | ~\mu \in \RR^\times \right\}$, so the
representation factors via $\RR^\times \times \PSL(\RR^2)$.  This
representation is conformal: the matrix $M_A$ preserves the lorentzian
inner product defined by
\begin{equation}
  K =
  \begin{pmatrix}
   \zero & 1 & \zero \\ 1 & \zero & \zero \\ \zero & \zero & 2
  \end{pmatrix};
\end{equation}
that is, $M_A^T K M_A = K$ and hence $\lambda^{-1} M_A^T K
\lambda^{-1} M_A = \lambda^{-2} K$.  Therefore $G$ is acting by
Lorentz transformations and an overall scale (which can be any non-zero
number).  The causal type of the vector is an invariant, but then by
using the rescaling we can bring the vector to one of four canonical
forms: the zero vector and a choice of spacelike, timelike
and null vector relative to $K$.  We may label these orbits by
choosing a representative vector $\t = (t_3, t_4, t_5)$ for each:
\begin{enumerate}
\item the \textbf{zero} orbit, where $\t = (0,0,0)$;
\item the \textbf{spacelike} orbit, where $\t = (0,0,1)$;
\item the \textbf{timelike} orbit, where $\t = (1,-1,0)$; and
\item the \textbf{lightlike} orbit, where $\t = (1,0,0)$.
\end{enumerate}
This gives four branches of solutions which we will study in turn.

\subsection{Zero branch deformations}
\label{sec:zero-branch-deform}

Here $t_3=t_4=t_5=0$, so that the system
\eqref{eq:quad-obstr} of quadrics becomes
\begin{equation}
  \begin{aligned}[m]
    t_1 t_2 &= 0 \\
    t_1 t_6 &= 0 \\
    t_2 t_6 &= 0 \\
  \end{aligned}
  \qquad\qquad
  \begin{aligned}[m]
    t_1 t_7 &= 0 \\
    t_1 t_8 &= 0 \\
    t_1 t_9 &= 0 \\
  \end{aligned}
  \qquad\qquad
  \begin{aligned}[m]
    t_2 t_7 &= 0 \\
    t_2 t_8 &= 0 \\
    t_2 t_9 &= 0 \\
  \end{aligned}
\end{equation}
and the system~\eqref{eq:cubic-obstr} of cubics is then identically
satisfied.  The deformation is given by
\begin{equation}
  \begin{split}
    \varphi_1 &= t_1 c_1 + t_2 (c_3 + c_7) + t_6 c_{10} +
    t_7(c_{12}-c_{16}) + t_8 (c_9 - c_{13}) + t_9 c_{15}\\
    \varphi_2 &= 2 (t_6 t_7 + t_8^2) c_8 + (t_7 t_8 + t_6 t_9) c_{11}
    + 2 (t_7^2 - t_8 t_9) c_{14}.
  \end{split}
\end{equation}

\subsubsection{$t_1 \neq 0$ subbranch}
\label{sec:t_1-neq-0}

If $t_1 \neq 0$, then $t_2 = t_6 = t_7 = t_8 = t_9 = 0$, so that
\begin{equation}
  \varphi_1 = t_1 c_1 \qquad\text{and}\qquad \varphi_2 = 0.
\end{equation}
Since $t_1 \neq 0$, we can change basis so that $t_1 = 1$ and end up
with the following Lie bracket (in addition to the ones involving
$R_i$):
\begin{equation}
  \label{eq:0-1-branch}
  \boxed{ [B_i, P_j] = \delta_{ij} H ,}
\end{equation}
which defines the \textbf{Carroll} algebra.

\subsubsection{$t_2 \neq 0$ subbranch}
\label{sec:t_2-neq-0}

If $t_2 \neq 0$, then $t_1 = t_6 = t_7 = t_8 = t_9 = 0$, so that
\begin{equation}
  \varphi_1 = t_2 (c_3 + c_7) \qquad\text{and}\qquad \varphi_2 = 0.
\end{equation}
Since $t_2 \neq 0$, we can change basis so that $t_2 = 1$ and end up
with
\begin{equation}
  \label{eq:eq:0-2-branch}
  \boxed{[H , B_i] = B_i \qquad\text{and}\qquad [H , P_i] = P_i.}
\end{equation}

\subsubsection{$t_1 = t_2 = 0$ subbranch}
\label{sec:t_1-=-t_2}

If $t_1 = t_2 = 0$, we are left with four parameters
$t_6,t_7,t_8,t_9$ defining the deformation
\begin{equation}
  \begin{split}
    \varphi_1 &= t_6 c_{10} + t_7(c_{12}-c_{16}) + t_8 (c_9 - c_{13}) + t_9 c_{15}\\
    \varphi_2 &= 2 (t_6 t_7 + t_8^2) c_8 + (t_7 t_8 + t_6 t_9) c_{11}
    + 2 (t_7^2 - t_8 t_9) c_{14}.
  \end{split}
\end{equation}
We still have the freedom to transform the parameters by the action of
$G$.  The four remaining parameters transform according to a
(conformally symplectic) rational representation of $\GL(\RR^2)$:
\begin{equation}
  \begin{pmatrix}
    t_6 \\ t_7 \\ t_8 \\ t_9
  \end{pmatrix} \mapsto
  \rho(A)   \begin{pmatrix}
    t_6 \\ t_7 \\ t_8 \\ t_9
  \end{pmatrix} := 
  \frac1{\Delta^2}
  \begin{pmatrix}
    d^3 & - 3 c^2 d & 3 c d^2 & c^3\\
    -b^2 d & a (ad + 2 bc) & -b (2 ad + bc) & -a^2 c\\
    bd^2 & -c(2ad+bc) & d(ad+2bc) & ac^2\\
    b^3 & - 3 a^2 b & 3 a b^2 & a^3
  \end{pmatrix}
  \begin{pmatrix}
    t_6 \\ t_7 \\ t_8 \\ t_9
  \end{pmatrix},
\end{equation}
where $\GL(\RR^2) \ni A =\begin{pmatrix} a & b \\ c & d \end{pmatrix}$
and $\Delta = \det A$.  The representation $A \mapsto \rho(A)$
satisfies
\begin{equation}
  \rho(A)^T \Omega \rho(A) = \frac1\Delta \Omega
  \qquad\text{where}\qquad \Omega =
  \begin{pmatrix}
    \zero & \zero & \zero & 1\\ \zero & \zero & -3 & \zero \\ \zero & 3 & \zero & \zero \\ -1 & \zero & \zero & \zero
  \end{pmatrix}.
\end{equation}
This representation is dual to the natural representation of
$\GL(\RR^2)$ on the space of binary cubics of the form
\begin{equation}
  t_9 X^3 - t_7 X^2 Y + t_8 XY^2 + t_6 Y^3,
\end{equation}
induced from the natural two-dimensional representation of
$\GL(\RR^2)$ on $(X,Y)$.  Classical invariant theory (see, e.g.,
\cite[p.28]{MR1694364}) tells us that there are four
$\GL(\RR^2)$-orbits in the space of (non-zero) real binary cubics,
characterised by  whether the corresponding binary cubics have three
distinct real roots, three distinct roots (only one real), a double
root or a triple root.  Choosing a representative from each orbit, we
have the following values of $\t = (t_6, t_7, t_8, t_9)$:

\begin{enumerate}
 
\item Three real roots: $\t = (0,0,1,1)$.  In this case, and after
  $\B\mapsto \tfrac13 (\B + \R)$ and $\P \mapsto \tfrac1{\sqrt3} \P$,
  we arrive at
  \begin{equation}
    \label{eq:three-real-roots-cubic}
    \boxed{[B_i, B_j] = \epsilon_{ijk} B_k \qquad\text{and}\qquad
      [P_i, P_j] = \epsilon_{ijk} (B_k - R_k).}
  \end{equation}

\item Three distinct roots, but only one real root:
  $\t = (0, 0, 1, -1)$ .  In this case, and after the same change of
  basis as in the previous case, we arrive at
  \begin{equation}
    \label{eq:three-not-real-roots-cubic}
    \boxed{[B_i, B_j] = \epsilon_{ijk} B_k \qquad\text{and}\qquad
      [P_i, P_j] = - \epsilon_{ijk} (B_k - R_k).}
  \end{equation}

\item A double root: $\t = (0,1,0,0)$.  In this case, and after
  $\P \mapsto -\tfrac13 (\P-\R)$, we arrive at
  \begin{equation}
    \label{eq:double-root-cubic}
    \boxed{[P_i, P_j] = \epsilon_{ijk} P_k.}
  \end{equation}

\item A triple root: $\t = (0,0,0,1)$.  This is simply
  \begin{equation}
    \label{eq:triple-root-cubic}
    \boxed{[P_i, P_j] = \epsilon_{ijk} B_k.}
  \end{equation}

\end{enumerate}

\subsection{Spacelike branch deformations}
\label{sec:spac-branch-deform}

Here $t_3 = t_4 = 0$ and $t_5 = 1$.  The deformations take the form
\begin{equation}
  \begin{split}
    \varphi_1 &= t_1 c_1 + (t_2 + 1) c_3 + (t_2 -1) c_7 + t_6 c_{10} + t_7
    (c_{12}-c_{16}) + t_8 (c_9-c_{13}) + t_9 c_{15}\\
    \varphi_2 &= 2 (t_6 t_7 + t_8^2) c_8 + (t_1 (t_2-1) + t_7 t_8 + t_6 t_9) c_{11} +
    2 ( t_7^2 - t_8 t_9) c_{14},
  \end{split}
\end{equation}
subject to the following systems of quadrics (the cubics become
quadrics in this case):
\begin{equation}
  \begin{aligned}[m]
    t_1 t_2 &= 0\\
    t_6 t_9 &= 0\\
    t_8 t_9 &= 0\\
  \end{aligned}
  \qquad\qquad
  \begin{aligned}[m]
    t_1 t_6 &= 0 \\
    t_1 t_7 &= 0\\
    t_1 t_8 &= 0\\
    t_1 t_9 &= 0
  \end{aligned}
  \qquad\qquad
  \begin{aligned}[m]
    (t_2+3)t_6 &= 0\\
    (t_2-1) t_7 &= 0\\
    (t_2+1) t_8 &= 0\\
    (t_2-3)t_9 &= 0
  \end{aligned}
\end{equation}
This system breaks up into several subbranches.

\subsubsection{$t_1\neq 0$ subbranch}
\label{sec:t_1neq-0-subbranch}

Here $t_1 \neq 0$, so that $t_2 = t_6= t_7= t_8 = t_9 = 0$.
The deformation has
\begin{equation}
  \varphi_1 = t_1 c_1 + c_3 - c_7 \qquad\text{and}\qquad \varphi_2 = - t_1 c_{11}
\end{equation}
leading (after rescaling of the generators) to
\begin{equation}
  \label{eq:1-1-branch}
  \boxed{[H ,B_i]=B_i \qquad [H ,P_i]=-P_i \qquad\text{and}\qquad
    [B_i,P_j] = \delta_{ij} H  - \epsilon_{ijk} R_k.}
\end{equation}
This Lie algebra is isomorphic to $\so(4,1)$, expressed relative to a
Witt (i.e., lightcone) basis, where $B_i$ plays the role of $L_{-i}$,
$P_i$ that of $L_{+i}$ and $H$ that of $L_{+-}$.  Since it is simple,
the Killing form is non-degenerate and hence it is a metric Lie
algebra.

\subsubsection{$t_1=0$, $t_2\neq \pm1,\pm 3$ subbranch}
\label{sec:t_1=0-t_2neq-0}

Here $t_1 = 0$ and $t_2 \neq \pm1,\pm3$, so that $t_6
= t_7 = t_8 = t_9=0$.  The deformation has $\varphi_1 = (t_2 + 1) c_3 + (t_2
- 1) c_7$, leading to
\begin{equation}
  [H ,B_i] = (t_2 +1) B_i \qquad\text{and}\qquad [H ,P_i] = (t_2 -1) P_i.
\end{equation}
The parameter $t_2$ can be further restricted by noticing that since
$t_2 \neq \pm 1$, we can rescale $H  \mapsto \tfrac1{t_2-1} H $, so
that
\begin{equation}
  \label{eq:1-2-branch}
  \boxed{[H , B_i] = \gamma B_i \qquad\text{and}\qquad [H , P_i] = P_i,}
\end{equation}
where we have introduced $\gamma := \tfrac{t_2+1}{t_2-1}$.  By
exchanging $\B \leftrightarrow \P$ if necessary, we can arrange so
that $\gamma \in [-1,1)$, the case $\gamma = -1$ corresponding to the
lorentzian \textbf{Newton} algebra.

\subsubsection{$t_1=0$ and $t_2 = 1$ subbranch}
\label{sec:t_1=0-t_2=1}

Here $t_1 = 0$ and $t_2 = 1$ and hence $t_6 = t_8 = t_9 = 0$.  The deformation has
\begin{equation}
  \varphi_1 = 2 c_3 + t_7 (c_{12} - c_{16}) \qquad\text{and}\qquad
  \varphi_2 = 2 t_7^2 c_{14}.
\end{equation}
There are two possible Lie algebras depending on whether or not $t_7=0$:
\begin{equation}
  \label{eq:1-3-1-branch}
  \boxed{[H ,B_i] = B_i,}
\end{equation}
which is the case $t_2 = 1$ of \eqref{eq:1-2-branch}, and
\begin{equation}
  \label{eq:1-3-2-branch}
  \boxed{[H ,B_i] = B_i \qquad\text{and}\qquad [P_i, P_j] =
    \epsilon_{ijk} P_k,}
\end{equation}
after redefining generators ($\P \mapsto \frac{-1}{3t_7}(\P - t_7 \R)$
and $H  \mapsto \tfrac12H$).

\subsubsection{$t_1 = 0$ and $t_2 = -1$ subbranch}
\label{sec:t_1-=0-t_2=-1}

Here $t_1 = 0$ and $t_2 =-1$, so that $t_6=t_7=t_9=0$.  This is
isomorphic to the previous case under $\B \leftrightarrow \P$.

\subsubsection{$t_1 = 0$ and $t_2=3$ subbranch}
\label{sec:t_1=-0-t_2=3}

Here $t_1 = 0$ and $t_2 = 3$, so that $t_6=t_7=t_8=0$.  The
deformation has $\varphi_1 = 4 c_3 + 2 c_7 + t_9 c_{15}$.  Rescaling
$H $ and, if $t_9\neq 0$, also $\B$ appropriately, we may bring the
Lie bracket to the following forms:
\begin{equation}
  \label{eq:1-6-1-branch}
  \boxed{[H ,B_i]=2B_i \qquad\text{and}\qquad [H ,P_i]=P_i,}
\end{equation}
if $t_9=0$, which is isomorphic to the case $t_2 = 3$ of
\eqref{eq:1-2-branch}, and
\begin{equation}
  \label{eq:1-6-2-branch}
  \boxed{[H ,B_i]=2B_i \qquad [H ,P_i]=P_i \qquad\text{and}\qquad
    [P_i,P_j]=\epsilon_{ijk} B_k,}
\end{equation}
if $t_9 \neq 0$.

\subsubsection{$t_1=0$ and $t_2 = -3$ subbranch}
\label{sec:t_1=0-t_2=-3}

This is equivalent to the previous case under $\B \leftrightarrow
\P$.

\subsection{Timelike branch deformations}
\label{sec:timel-branch-deform}

Here $t_5=0$, $t_3 = 1$ and $t_4 = -1$.  The system
\eqref{eq:quad-obstr} of quadrics becomes
\begin{equation}
  \begin{aligned}[m]
    t_1 t_2 & = 0\\
    t_1 t_6 & = 0\\
    t_1 t_7 & = 0\\
    t_1 t_8 &= 0\\
    t_1 t_9 &= 0
  \end{aligned}
  \qquad\qquad
  \begin{aligned}[m]
    t_2 t_6 &= t_8\\
    t_2 t_7 &= 2 t_8 - t_9\\
    t_2 t_8 &= -\tfrac35 t_6 - \tfrac65 t_7\\
    t_2 t_9 &= 3 t_7.\\
  \end{aligned}
\end{equation}
The four equations on the right can be written in the following
suggestive form:
\begin{equation}
  \begin{pmatrix}
    \zero & \zero & 1 & \zero\\
    \zero & \zero & 2 & -1\\
    -\tfrac35 & -\tfrac65 & \zero & \zero\\
    \zero & 3 & \zero & \zero
  \end{pmatrix}
  \begin{pmatrix}
    t_6 \\ t_7 \\ t_8 \\t_9
  \end{pmatrix}
  = t_2 \begin{pmatrix}
    t_6 \\ t_7 \\ t_8 \\t_9
  \end{pmatrix},
\end{equation}
which says that $(t_6,t_7,t_8,t_9)$ is an eigenvector of the matrix on
the left-hand side with (real) eigenvalue $t_2$ and therefore also an eigenvector
of the square of that matrix with non-negative eigenvalue $t_2^2$:
\begin{equation}
  \begin{pmatrix}
    -\tfrac35 & -\tfrac65 & \zero & \zero \\[3pt]
    -\tfrac65 & -\tfrac{27}5 & \zero & \zero \\
    \zero & \zero & -3 & \tfrac65\\
    \zero & \zero & 6 & -3
  \end{pmatrix}
  \begin{pmatrix}
    t_6 \\ t_7 \\ t_8 \\t_9
  \end{pmatrix} =
  t_2^2 \begin{pmatrix}
    t_6 \\ t_7 \\ t_8 \\t_9
  \end{pmatrix}.
\end{equation}
Notice, however, that the above matrix is diagonalisable (over the reals)
with \emph{negative} eigenvalues $-\tfrac35 ( 5 \pm 2 \sqrt5)$, each
with multiplicity $2$.  Therefore the above equation has as unique solution $t_6
= t_7 = t_8 = t_9 = 0$, which automatically solves the cubic
system~\eqref{eq:cubic-obstr} and leaves the following $t_1 t_2 = 0$,
$t_3 = 1$ and $t_4 = -1$.  This gives rise to two
branches depending on whether or not $t_1 = 0$.

\subsubsection{$t_1 \neq 0$ subbranch}
\label{sec:t_1-neq-0-1}

If $t_1 \neq 0$, then $t_2 = 0$ and the deformation has
\begin{equation}
  \varphi_1 = t_1 c_1 + c_4 - c_6 \qquad\text{and}\qquad \varphi_2 =
  t_1 c_8 + t_1 c_{14},
\end{equation}
which results in the following Lie brackets:
\begin{equation}
  \begin{aligned}[m]
    [H  , B_i ] &=  P_i\\
    [H  , P_i ] &= - B_i\\
    [B_i , P_j ] &= t_1 \delta_{ij} H 
  \end{aligned}
  \qquad\qquad
  \begin{aligned}[m]
    [B_i , B_j ] &= t_1 \epsilon_{ijk} R_k\\
    [P_i , P_j ] &= t_1 \epsilon_{ijk} R_k\\
  \end{aligned}
\end{equation}
We may rescale $\B$ and $\P$ to set $t_1 = \pm 1$, depending on its
sign, and also rescale $H$ by that sign leading to the Lie algebra
\begin{equation}
  \label{eq:2-1-branch}
  \boxed{
    \begin{aligned}[m]
      [B_i,P_j] &= \delta_{ij} H \\
      [H ,B_i] &= \pm P_i\\
      [H ,P_i] &= \mp B_i\\
    \end{aligned}
    \qquad\qquad
    \begin{aligned}[m]
      [B_i,B_j] &= \pm \epsilon_{ijk} R_k\\
      [P_i,P_j] &= \pm \epsilon_{ijk} R_k,
    \end{aligned}
  }
\end{equation}
which is isomorphic either to $\so(5)$ or to $\so(3,2)$, depending on
the sign.  Since these Lie algebras are simple, they are metric
relative to the Killing form.

\subsubsection{$t_1 = 0$ subbranch}
\label{sec:t_1-=-0-2}

Here the deformation is
\begin{equation}
  \varphi_1 = t_2 (c_3 + c_7) + c_4 - c_6
\end{equation}
which leads to the following Lie algebra
\begin{equation}
  \label{eq:2-2-branch}
  \boxed{ [ H ,B_i] = t_2 B_i + P_i \qquad\text{and}\qquad [H ,P_i]
    = t_2 P_i - B_i,}
\end{equation}
where we can always arrange $t_2\geq 0$ by relabelling generators.  If
$t_2 = 0$ this is the euclidean \textbf{Newton} algebra.

\subsection{Lightlike branch deformations}
\label{sec:lightl-branch-deform}

Here $t_3 = 1$ and $t_4 = t_5 = 0$.  The system \eqref{eq:quad-obstr}
of quadrics becomes
\begin{equation}
  \begin{aligned}[m]
    t_1 t_2 &= 0\\
    t_1 t_6 &=0\\
    t_1 t_7 &=0\\
  \end{aligned}
  \qquad\qquad
  \begin{aligned}[m]
    t_1 t_8 &=0\\
    t_1 t_9 &=0\\
    t_2 t_9 &= 0 \\
  \end{aligned}
  \qquad\qquad
  \begin{aligned}[m]
    t_7 &= -\tfrac56 t_2 t_8\\
    t_8 &=  t_2 t_6\\
    t_9 &= -t_2 t_7.\\
  \end{aligned}
\end{equation}
Plugging $t_9 = -t_2 t_7$ into $t_2 t_9 = 0$, yields $t_2^2 t_7 = 0$
which implies that $t_2 t_7 = 0$, so that $t_9 = 0$.  Similarly, $t_2
t_7 = 0$ implies that $t_2 t_8 = 0$ and hence that $t_7 = 0$, and
finally this implies that $t_8 = 0$ as well.  This already means that
the cubic system~\eqref{eq:cubic-obstr} is identically satisfied and
when the dust clears we are left with $t_3 = 1$, $t_4 = t_5 = t_7 =
t_8 = t_9= 0$, subject to the following conditions:
\begin{equation}
  t_1 t_2 = 0 \qquad t_1 t_6 = 0 \qquad\text{and}\qquad t_2 t_6 = 0.
\end{equation}
This gives rise to three branches of solutions.

\subsubsection{$t_1\neq 0$ subbranch}
\label{sec:t_1neq-0-subbranch-1}

In this case, $t_1\neq 0$ and hence $t_2=t_6=0$.  The deformation is
\begin{equation}
  \varphi_1 = t_1 c_1 + c_4 \qquad\text{and}\qquad \varphi_2 = t_1 c_8,
\end{equation}
resulting in the following Lie brackets:
\begin{equation}
  [H ,B_i] = P_i \qquad [B_i, P_j] = t_1 \delta_{ij} H 
  \qquad\text{and}\qquad [B_i, B_j] = t_1 \epsilon_{ijk} R_k.
\end{equation}
We may rescale the generators in such a way that we reabsorb $t_1$ up
to its sign and arrive at two non-isomorphic deformations
\begin{equation}
  \label{eq:3-1-branch}
  \boxed{[H ,B_i] = \pm P_i \qquad [B_i,P_j] = \delta_{ij} H 
    \qquad\text{and}\qquad [B_i, B_j] = \pm \epsilon_{ijk} R_k,}
\end{equation}
which correspond to the \textbf{euclidean} $\e$ and \textbf{Poincaré}
$\p$ Lie algebras.

\subsubsection{$t_2\neq 0$ subbranch}
\label{sec:t_2neq-0-subbranch}

Here $t_2\neq 0$, so $t_1 = t_6 = 0$.  The deformation is therefore
\begin{equation}
  \varphi_1 = t_2 (c_3 + c_7) + c_4,
\end{equation}
which after rescaling generators can be brought to the form
\begin{equation}
  \label{eq:3-2-branch}
  \boxed{[H ,B_i] = B_i + P_i \qquad\text{and}\qquad [H , P_i] = P_i.}
\end{equation}

\subsubsection{$t_1 = t_2 = 0$ subbranch}
\label{sec:t_1-=-t_2-1}

Here $t_1 = t_2 = 0$ and hence the deformation is given by
\begin{equation}
  \varphi_1 = c_4 + t_6 c_{10},
\end{equation}
leading to the Lie brackets
\begin{equation}
  [H , B_i] = P_i \qquad\text{and}\qquad [B_i,B_j] = t_6
  \epsilon_{ijk} P_k.
\end{equation}
If $t_6 = 0$, we arrive at the \textbf{galilean} algebra (after rescaling)
\begin{equation}
  \label{eq:galilean}
  \boxed{[H , B_i] = - P_i.}
\end{equation}
If $t_6 \neq 0$, we can rescale the generators to arrive at
\begin{equation}
  \label{eq:3-3-branch}
  \boxed{[H ,B_i] = - P_i \qquad\text{and}\qquad [B_i, B_j] =
    \epsilon_{ijk} P_k.}
\end{equation}

\subsection{Invariant inner products}
\label{sec:invar-inner-prod}

We shall now analyse the existence of invariant inner products on the
Lie algebras determined in this section.  Recall that an invariant
inner product on a Lie algebra $\g$ is a non-degenerate symmetric
bilinear form $(-,-): \g \times \g \to \RR$ which is ``associative'';
that is,
\begin{equation}\label{eq:assoc}
  ([x,y],z) = (x,[y,z]) \qquad\text{for all~} x,y,z \in \g.
\end{equation}
The Killing form is associative, but it is only non-degenerate for
semisimple Lie algebras, so that the inner product on non-semisimple
metric Lie algebras is always an additional piece of data.  When it
exists, it is seldom positive-definite, unless $\g$ is the Lie algebra
of a compact group. This means that it is the direct sum of a
semisimple Lie algebra (of compact type) and an abelian Lie algebra.

Rather than appealing to any general structural results, the strategy
here is simply to exploit the associativity condition
\eqref{eq:assoc}.  We shall first of all show that no kinematical
Lie algebra where $\B$ and $\P$ span an abelian ideal can be metric.
This will rule out the first eight cases in
Table~\ref{tab:kla-summary}.  Indeed, let $(-,-)$ be an associative
symmetric bilinear form.  We will show that it is degenerate.  To this
end, let $X,Y$ be any of $B,P$ and consider
\begin{equation}
  \epsilon_{ijk} (X_k, Y_\ell) = ([R_i,X_j],Y_\ell) = (R_i, [X_j,
  Y_\ell]) = 0,
\end{equation}
where we have used associativity and the fact that $X,Y$ are vectors
under rotations.  Therefore the only non-zero components of $(-,-)$ are
\begin{equation}
  (H,H) \qquad (R_i,R_j) \qquad (R_i,B_j) \qquad (R_i,P_j)
\end{equation}
and hence there is some non-zero $Z_i = \alpha B_i + \beta P_i$, for
some $\alpha,\beta \in \RR$ (not both zero), which obeys $(Z_i,-) =
0$.

Any associative symmetric bilinear form in the Carroll algebra is
degenerate, since $(H,-) = 0$.  Indeed, by rotational invariance, the
only possible non-zero inner product of $H$ is with itself, but then
\begin{equation}
  \delta_{ij} (H,H) = (H, [B_i, P_j]) = ([H,B_i],P_j) = 0.
\end{equation}

The simple Lie algebras $\so(4,1)$, $\so(5)$ and $\so(3,2)$ are of
course metric relative to the Killing form, whereas the euclidean and
Poincaré algebras (in this dimension) are not metric.  Indeed, let
$(-,-)$ be an associative symmetric bilinear form on either $\e$ or
$\p$ and calculate $(H,H)$, which is the only possibly non-zero
rotationally invariant inner product involving $H$:
\begin{equation}
  \delta_{ij} (H,H) = (H, [B_i, P_j]) = - (H, [P_j, B_i]) = -
  ([H,P_j], B_i) = 0.
\end{equation}

This settles all the Lie algebras above the line in
Table~\ref{tab:kla-summary}.  Of the seven Lie algebra below the line
in that table, it will turn out that the first four are metric, but
not the last three.  Let's do them first.

Consider the Lie algebra in \eqref{eq:1-3-2-branch} and let $(-,-)$ be
an associative symmetric bilinear form.  If $X$ is any one of $R,B$,
then
\begin{equation}
  (P_i, X_j) = ([H,P_i],X_j) = - ([P_i,H], X_j) = - (P_i,[H,X_j]) = 0,
\end{equation}
whereas
\begin{equation}
  (P_i, P_j) = ([H,P_i],P_j) = (H,[P_i,P_j]) = 0.
\end{equation}
Therefore, $(P_i,-) = 0$.

Let $(-,-)$ be an associative symmetric bilinear form on the Lie
algebra in \eqref{eq:3-3-branch}.  Then again if $X$ is any of $R,P$,
\begin{equation}
  (P_i,X_j) = ([B_i,H],X_j) = (B_i, [H,X_j]) = 0,
\end{equation}
whereas
\begin{equation}
  (B_i,P_j) = (B_i, [B_j,H]) = ([B_i,B_j],H) = \epsilon_{ijk} (P_k, H)
  = 0,
\end{equation}
by rotational invariance.  Therefore $(P_i,-) = 0$.

Let $(-,-)$ be an associative symmetric bilinear form on the Lie
algebra in \eqref{eq:1-6-2-branch}.  Then if $X$ is either $B$ or $P$,
\begin{equation}
  (B_i, X_j) = \tfrac12 ([H,B_i],X_j) = \tfrac12 (H,[B_i,X_j]) = 0,
\end{equation}
whereas
\begin{equation}
  (R_i, B_j) = \tfrac12 (R_i, [H,B_j]) = \tfrac12 ([R_i,H], B_j) = 0,
\end{equation}
so that $(B_i,-) = 0$.

The first four Lie algebras below the line in
Table~\ref{tab:kla-summary} are metric under a four-parameter family
of associative inner products.  For these algebras $H$ remains
central, so one of the parameters is $(H,H)$, which has to be
different from zero.  To describe the other three parameters, let us
encode the associative inner product on the nine-dimensional
subalgebra spanned by $R_i,B_i,P_i$ as a $3 \times 3$ symmetric
matrix:
\begin{equation}
  \begin{pmatrix}
    b_{11} & b_{12} & b_{13} \\ b_{12} & b_{22} & b_{23} \\ b_{13} &
    b_{23} & b_{33}
  \end{pmatrix}
  \quad\text{where}\quad (R_i,R_j) = b_{11} \delta_{ij}, (R_i,B_j) =
  b_{12} \delta_{ij}, \dots, (P_i,P_j) = b_{33} \delta_{ij};
\end{equation}
although it is important to keep in mind that the non-degeneracy of the
inner product is \emph{not equivalent to} the non-degeneracy of this
symmetric matrix.  (It is the trace, not the determinant, which is
multiplicative over the tensor product.)  We will simply list the
matrices for each of the Lie algebras in question, along with the
condition of non-degeneracy on the parameters.

For the Lie algebra in \eqref{eq:three-real-roots-cubic}, we have
\begin{equation}
  \begin{pmatrix}
    \alpha & \beta & \gamma \\ \beta & \alpha & \zero \\ \gamma & \zero &
    \beta - \alpha    
  \end{pmatrix} \qquad \beta ((\alpha-\beta)^2 + \gamma^2) \neq 0.
\end{equation}
For the Lie algebra in \eqref{eq:three-not-real-roots-cubic}, we have
\begin{equation}
  \begin{pmatrix}
    \alpha & \beta & \gamma \\ \beta & \alpha & \zero \\ \gamma & \zero &
    \alpha - \beta    
  \end{pmatrix} \qquad \beta ((\alpha-\beta)^2 + \gamma^2) \neq 0.
\end{equation}
For the Lie algebra in \eqref{eq:double-root-cubic}, we have
\begin{equation}
  \begin{pmatrix}
    \alpha & \beta & \gamma \\ \beta & \beta & \zero \\ \gamma & \zero &
    \zero
  \end{pmatrix} \qquad \beta \gamma \neq 0.
\end{equation}
Finally, for the Lie algebra in \eqref{eq:triple-root-cubic}, we have
\begin{equation}
  \begin{pmatrix}
    \alpha & \beta & \gamma \\ \beta & \gamma & \zero \\  \gamma & \zero &
    \zero
  \end{pmatrix} \qquad \gamma \neq 0.
\end{equation}

\subsection{Summary}
\label{sec:summary}

The classification in this section is of course not new: the Lie
algebras agree precisely with the kinematical Lie algebras classified
by Bacry and Nuyts in \cite{MR857383}.  Table~\ref{tab:kla-summary}
lists our results and they can be compared with Table~1 in that paper.
Our notation differs from that in \cite{MR857383} in that our $\R$ and
$\B$ are their $\J$ and $\K$, respectively.

All kinematical Lie algebras share the following Lie brackets (in
abbreviated notation):
\begin{equation}
  [\R,\R] = \R \qquad [\R,\B] = \B \qquad [\R, H ] = 0
  \qquad\text{and}\qquad [\R, \P] = \P,
\end{equation}
so in the table we will only list any additional brackets.  The
\textbf{static} kinematical Lie algebra has no additional non-zero
brackets and is listed first, for completeness.  In some cases we have
relabelled generators ($\B \leftrightarrow \P$) in order to arrive at
a more uniform description.  It follows from the classifications of
kinematical Lie algebras in dimension $D+1$ for $D\geq 4$
\cite{JMFKinematicalHD} and for $D=2$ \cite{TAJMFKinematical2D} that
the kinematical Lie algebras in Table~\ref{tab:kla-summary} which lie
below the line are unique to $D=3$: indeed, they owe their existence
to the vector product in $\RR^3$, which is invariant under rotations.

\begin{table}[h!]\tiny
  \centering
  \caption{Kinematical Lie algebras}
  \label{tab:kla-summary}
  \setlength{\extrarowheight}{2pt}  
  \begin{tabular}{l|*{5}{>{$}l<{$}}|l|c}
    \multicolumn{1}{c|}{Eq.} & \multicolumn{5}{c|}{Non-zero Lie brackets} & \multicolumn{1}{c|}{Comments} & \multicolumn{1}{c}{Metric?}\\\hline
    \ref{eq:static} & & & & & & static &  \\
    \ref{eq:galilean} & [H ,\B] = -\P & & & & & galilean &  \\
    \ref{eq:1-2-branch} & [H ,\B] = \gamma \B & [H ,\P] = \P & & & & $0,\tfrac12 \neq \gamma \in (-1,1)$ &  \\
    \ref{eq:1-2-branch} & [H ,\B] = - \B & [H ,\P] = \P & & & & lorentzian Newton &  \\    
    \ref{eq:eq:0-2-branch} & [H ,\B] = \B & [H ,\P] = \P & & & & $\gamma=1$ in \eqref{eq:1-2-branch} &  \\
    \ref{eq:1-3-1-branch} & [H ,\B] = \B & & & & & $\gamma = 0$ in \eqref{eq:1-2-branch} &  \\
    \ref{eq:1-6-1-branch} & [H ,\B] = 2\B & [H ,\P] = \P & & & & $\gamma=\tfrac12$ in \eqref{eq:1-2-branch} &  \\
    \ref{eq:2-2-branch} & [H ,\B] = \alpha \B + \P & [H ,\P] = \alpha \P - \B & & & & $\alpha > 0$ &  \\
    \ref{eq:2-2-branch} & [H ,\B] = \P & [H ,\P] = - \B & & & & euclidean Newton &  \\
    \ref{eq:3-2-branch} & [H ,\B] = \B + \P & [H , \P] = \P & & & &  &  \\
    \ref{eq:0-1-branch} & & & [\B,\P] = H  & & & Carroll &  \\
    \ref{eq:3-1-branch} & [H ,\B] = \P & & [\B,\P] = H  & [\B,\B] = \R & & $\e$ (euclidean) &  \\
    \ref{eq:3-1-branch} & [H ,\B] = - \P & & [\B,\P] = H  & [\B,\B] = - \R & & $\p$ (Poincaré) &  \\
    \ref{eq:1-1-branch} & [H ,\B] = \B & [H ,\P] = -\P &  [\B,\P] = H  - \R & & & $\so(4,1)$ & \checkmark \\
    \ref{eq:2-1-branch} & [H ,\B] = \P & [H ,\P] = -\B & [\B,\P] = H  &  [\B,\B]= \R &  [\P,\P] = \R & $\so(5)$ & \checkmark\\
    \ref{eq:2-1-branch} & [H ,\B] = -\P & [H ,\P] = \B & [\B,\P] = H  &  [\B,\B]= -\R &  [\P,\P] = -\R & $\so(3,2)$ & \checkmark\\\hline
    \ref{eq:three-real-roots-cubic} & & & & [\B,\B]= \B &  [\P,\P] = \B-\R & & \checkmark\\
    \ref{eq:three-not-real-roots-cubic} & & & & [\B,\B]= \B & [\P,\P] = \R-\B & & \checkmark\\
    \ref{eq:double-root-cubic} & & & & [\B,\B] = \B & &  & \checkmark\\
    \ref{eq:triple-root-cubic} & & & & [\B, \B] = \P & &  & \checkmark\\
    \ref{eq:1-3-2-branch} & & [H ,\P] = \P & & [\B,\B] = \B & &  &  \\
    \ref{eq:3-3-branch} & [H ,\B] = -\P & & & [\B,\B] = \P & &  &  \\
    \ref{eq:1-6-2-branch} & [H ,\B] = \B & [H ,\P] = 2\P & & [\B,\B] = \P & & &  \\
  \end{tabular}
\end{table}

\section{Deformations of the centrally-extended static kinematical Lie
  algebra}
\label{sec:deform-centr-extend}

As shown in Appendix~\ref{sec:centr-extens-stat}, the static
kinematical Lie algebra $\g$ given by \eqref{eq:static} admits a
one-dimensional universal central extension $\tg$, generated by $R_i,
B_i, P_i, H , Z$ and non-zero Lie brackets in abbreviated notation:
\begin{equation}
  \label{eq:static-centrext}
  [\R,\R] = \R \qquad [\R,\B] = \B \qquad [\R,\P] = \P
  \qquad\text{and}\qquad [\B,\P] = Z.
\end{equation}
We will let $\th$ denote the ideal generated by $\B,\P,H ,Z$ and
again $\s$ the rotational subalgebra generated by $\R$.  By the
Hochschild--Serre decomposition theorem, $H^2(\tg;\tg) \cong
H^2(\th;\tg)^{\s}$, which can be calculated by the $\s$-invariant
subcomplex $C^\bullet(\th;\tg)^{\s}$ described in
Appendix~\ref{sec:coch-centr-stat-kinem}.  As in the case of the
static kinematical Lie algebra $\g$ treated in
Section~\ref{sec:deform-stat-kinem}, it will be convenient to exploit
the action of those automorphisms of $\th$ which commute with $\s$.

\subsection{Automorphisms of $\th$}
\label{sec:automorphisms-hb}

Let $\tG = \GL(\RR^2) \ltimes \Aff(\Lambda^2\RR^2)$ denote the
semidirect product of $\GL(\RR^2)$, the group of invertible linear
transformations of $\RR^2$, and $\Aff(\Lambda^2\RR^2)$, the group of
invertible affine transformations on the one-dimensional vector space
$\Lambda^2\RR^2$.  The reason we do not simply call this $\RR$ is that
$\Lambda^2\RR^2$ is the one-dimensional (determinant) representation
of $\GL(\RR^2)$.  This group acts on $\th$ by automorphisms as follows:
\begin{equation}
  (\B,\P,H ,Z) \mapsto (\B, \P, H ,Z)
  \begin{pmatrix} 
    a & b & \zero & \zero \\ c & d & \zero & \zero \\ \zero & \zero & \lambda & \zero \\ \zero & \zero &
    \mu \Delta & \Delta
  \end{pmatrix} = (a \B + c \P, b \B + d \P, \lambda H  + \mu \Delta
  Z, \Delta Z),
\end{equation}
where $\Delta = \det A = ad - bc $.  The induced action on $\th^*$ is
\begin{equation}
  \bbeta \mapsto \Delta^{-1}(d \bbeta - b \bpi) \qquad \bpi \mapsto
  \Delta^{-1}(-c \bbeta + a \bpi) \qquad \eta  \mapsto \lambda^{-1}
  \eta  \qquad\text{and}\qquad \zeta \mapsto \Delta^{-1} \zeta -
  \lambda^{-1}\mu \eta .
\end{equation}

\subsection{Infinitesimal deformations}
\label{sec:infin-deform-1}

In the notation of Appendix~\ref{sec:coch-centr-stat-kinem} and taking
into account the action of the Chevalley--Eilenberg differential on
$\s$-invariant cochains given by
Equation~\eqref{eq:CE-static-ce-explicit} we see that the spaces of
coboundaries $B^2$ and cocycles $Z^2$ are given by
\begin{equation}
  B^2 = \RR\left<\ct_1,\ct_2,\ct_4+2\ct_{11},\ct_5+2\ct_7\right>
\end{equation}
and
\begin{equation}
  Z^2 = B^2 \oplus \RR\left<\ct_{14} + \ct_{16} + 2 \ct_{24}, \ct_{14}
    - \ct_{16}, \ct_{13}, \ct_3 + \ct_{20} - \ct_{22}, \ct_{17},
    \ct_9-\ct_{19},\ct_6+\ct_{23}\right>,
\end{equation}
where the chosen basis is adapted to the $\tG$-action.  The most
general (non-trivial) infinitesimal deformation is parametrised by
$(u_1,\dots,u_7) \in \RR^7$ as
\begin{multline}
  \varphi_1 = u_1 (\ct_{14} + \ct_{16} + 2 \ct_{24}) - u_2 (\ct_{14}
  - \ct_{16}) + u_3 \ct_{13} - u_4 \ct_{17}\\
  + u_5 (\ct_3 + \ct_{20} - \ct_{22}) + u_6 (\ct_9-\ct_{19}) + u_7
  (\ct_6+\ct_{23}),
\end{multline}
where we have altered the signs of $u_2$ and $u_4$ to obtain cleaner
formulae later on.

\subsection{Obstructions}
\label{sec:obstructions-1}

The first obstruction is the class of $\tfrac12
[\![\varphi_1,\varphi_1]\!]$.  Using the
formulae~\eqref{eq:NR-ext-static} for the restriction of the
Nijenhuis--Richardson bracket to the $\s$-invariant cochains, we find
that
\begin{multline}
  \tfrac12 [\![\varphi_1,\varphi_1]\!] = (2u_1 u_5 + u_3 u_7 - u_4
  u_6) (\bt_6 + \bt_{22} - \bt_{2})\\
  + (u_1 u_7 + u_2 u_7 - u_4 u_5) (2 \bt_3 + \bt_{19}) + (u_3 u_5 -
  u_1 u_6 + u_2 u_6) (\bt_{25} - 2 \bt_5).
\end{multline}
These cocycles are all non-trivial and linearly independent in
cohomology, so this obstruction vanishes if and only if the cocycle
vanishes.  This means that the integrability locus is the solution of
the system of quadrics
\begin{equation}
  \label{eq:quadrics-cext}
  \begin{split}
    2u_1 u_5 + u_3 u_7 - u_4 u_6 &= 0\\
    u_1 u_7 + u_2 u_7 - u_4 u_5 &= 0 \\
    u_3 u_5 - u_1 u_6 + u_2 u_6 &= 0.
  \end{split}
\end{equation}
If these equations are satisfied, $[\![\varphi_1,\varphi_1]\!] = 0$ so
that we can take $\varphi_2 = 0$ and hence there are no further
obstructions.  We study the system~\eqref{eq:quadrics-cext} by first
exploiting the action of the automorphisms in order to bring the
parameters to normal forms.

\subsection{The action of automorphisms on the deformation parameters}
\label{sec:action-automorphisms}

The action of $\tG$ on the cochains induces a linear action on the
parameter space, which can be described as follows:
\begin{equation}
  \label{eq:action-tG}
  \u = (u_1,\dots,u_7)^T \mapsto
  \begin{pmatrix}
    \lambda^{-1} & 0 & 0 \\
    0 & \lambda^{-1} \rho(A) & \lambda^{-1} \mu \rho(A)\\
    0 & 0 & \Delta^{-1} \rho(A)
  \end{pmatrix}
  \u,
\end{equation}
where $\lambda \in \RR^\times$, $\mu \in \RR$ and if $A = \begin{pmatrix} a & b \\ c & d \end{pmatrix} \in
\GL(\RR^2)$, $\Delta = \det A = a d - b c$ and
\begin{equation}
  \rho(A) =\frac1\Delta
  \begin{pmatrix}
    a d + b c & - a c & - b d \\ -2 a b & a^2 & b^2 \\ - 2 c d & c^2 &
    d^2
  \end{pmatrix}.
\end{equation}
The representation $\rho$ of $\GL(\RR^2)$ defined by
$A \mapsto \rho(A)$ has kernel $\{a \id \mid a \in \RR^\times\}$, the
group of scalar matrices, so that it descends to a representation of
the projective linear group $\PSL(\RR^2) \cong \SO(2,1)_o$, the
identity component of the three-dimensional Lorentz group.  Indeed,
this representation preserves a lorentzian inner product on the
three-dimensional space of parameters $(u_5, u_6, u_7)$:
\begin{equation}
  \rho(A)^T K \rho(A) = K \qquad\text{for}\qquad K =
  \begin{pmatrix}
    2 & \zero & \zero \\ \zero & \zero & -1 \\ \zero & -1 & \zero
  \end{pmatrix}.
\end{equation}
Since we only have at our disposal the identity component of the
Lorentz group, we preserve time-orientation for causal vectors.
Therefore we have the following six normal forms, with the
corresponding $\t = (u_5, u_6, u_7)$:
\begin{enumerate}
\item the \textbf{zero} orbit, where $\t = (0,0,0)$;
\item the \textbf{spacelike} orbit, where $\t = (1,0,0)$;
\item the \textbf{future timelike} orbit, where $\t = (0,1,1)$;
\item the \textbf{past timelike} orbit, where $\t = (0,-1,-1)$;
\item the \textbf{future lightlike} orbit, where $\t = (0,0,1)$; and
\item the \textbf{past lightlike} orbit, where $\t = (0,0,-1)$.
\end{enumerate}
This gives six branches of solutions which we will study in turn.

\subsection{Zero branch deformations}
\label{sec:zero-branch-deform-1}

In this case $u_5=u_6=u_7=0$ and the infinitesimal deformation is
already integrated:
\begin{equation}
  \varphi_1 = u_1 (\ct_{14} + \ct_{16} + 2 \ct_{24}) - u_2 (\ct_{14}
  - \ct_{16}) + u_3 \ct_{13} - u_4 \ct_{17}.
\end{equation}
The additional Lie brackets are (in abbreviated form)
\begin{equation}
  \begin{split}
    [H ,\B] &= (u_1 + u_2) \B - u_4 \P\\
    [H ,\P] &= u_3 \B + (u_1 - u_2) \P\\
    [H ,Z] &= 2 u_1 Z.
  \end{split}
\end{equation}
We must distinguish two subbranches, depending on whether or not $u_1
= 0$.

\subsubsection{$u_1 = 0$ subbranch}
\label{sec:u_1-=-0}

If $u_1 = 0$, we obtain 
\begin{equation}
  \begin{split}
    [H ,\B] &= u_2 \B - u_4 \P\\
    [H ,\P] &= u_3 \B - u_2 \P,
  \end{split}
\end{equation}
which, depending on the sign of the discriminant
$\delta := u_2^2 - u_3u_4$, is isomorphic to a (non-trivial) central
extension of one of the following deformations of the static
kinematical Lie algebra:
\begin{enumerate}
\item $\delta > 0$: then we can change basis so that
  \begin{equation}\label{eq:celornewton}
    \boxed{[H , \B] = \B \qquad\text{and}\qquad [H ,\P] = -\P,}
  \end{equation}
  which is isomorphic to the lorentzian Newton algebra.  The
  corresponding deformation is the well-known universal central
  extension of the lorentzian Newton algebra.

\item $\delta < 0$: then we can change basis so that
  \begin{equation}\label{eq:ceeucnewton}
    \boxed{[H , \B] = \P \qquad\text{and}\qquad [H ,\P] = -\B,}
  \end{equation}
  which is isomorphic to the euclidean Newton algebra.  The
  corresponding deformation is now the well-known universal central
  extension of the euclidean Newton algebra.

\item $\delta = 0$: then we can change basis so that
  \begin{equation}\label{eq:bargmann}
    \boxed{[H ,\B] = -\P,}
  \end{equation}
  isomorphic to the galilean algebra.  In other words, this
  deformation is isomorphic to the \textbf{Bargmann algebra}: the
  universal central extension of the galilean algebra.
\end{enumerate}

\subsubsection{$u_1 \neq 0$ subbranch}
\label{sec:u_1-neq-0}

If $u_1 \neq 0$ then we obtain a \emph{non-central} extension of some
of the deformations of the static kinematical Lie algebra.  Indeed,
$Z$ generates an ideal and quotienting by this ideal gives, depending
on the values of $(u_1, u_2, u_3, u_4)$, one of the deformations of
the static kinematical Lie algebra.  The action of $\tG$ on
the subspace of parameters with $u_5=u_6=u_7=0$ can be read off from
Equation~\eqref{eq:action-tG}:
\begin{equation}
  \begin{pmatrix}
    u_1\\u_2\\u_3\\u_4
  \end{pmatrix} \mapsto
  \frac1{\lambda \Delta}
  \begin{pmatrix}
    \Delta & \zero & \zero & \zero \\
    \zero & a d + b c & - a c & - b d \\ \zero & -2 a b & a^2 & b^2 \\ \zero & - 2 c d &
    c^2 & d^2
  \end{pmatrix}
  \begin{pmatrix}
    u_1\\u_2\\u_3\\u_4
  \end{pmatrix}.
\end{equation}
Taking $\lambda = u_1$, we can set $u_1=1$ without loss of
generality.  The remaining parameters transform under $\GL(\RR^2)$ as
a three-dimensional vector under the identity component of the Lorentz
group.  The orbits are classified by their lorentzian norm
$u_2^2-u_3u_4$, which can be any real number.  We obtain therefore the
following isomorphism classes of deformations:
\begin{enumerate}
\item $u_2^2 - u_2 u_4 > 0$:
  \begin{equation}
    \label{eq:zero-spacelike}
    \boxed{[H ,\B] = \gamma \B \qquad [H ,\P] = \P \qquad [H , Z] =
      (\gamma + 1) Z,}
  \end{equation}
  for $\gamma \in (-1,1]$.  This Lie algebra $\hat\g$ is a non-central
  extension
  \begin{equation}
    \begin{CD}
      0 @>>> \RR\left<Z\right> @>>> \hat\g @>>> \hat\h @>>> 0,
    \end{CD}
  \end{equation}
  of the Lie algebra, denoted here by $\hat h$, given by equations
  \eqref{eq:1-2-branch} (for $\gamma \neq -1,0,\tfrac12,1$),
  \eqref{eq:eq:0-2-branch} (for $\gamma=1$), \eqref{eq:1-3-1-branch}
  (for $\gamma = 0$) or \eqref{eq:1-6-1-branch} (for
  $\gamma=\tfrac12$.)  The limiting case $\gamma = -1$ is the central
  extension of the lorentzian Newton algebra discussed above
  (corresponding to $u_1 = 0$).

\item $u_2^2 - u_2 u_4 = 0$:
  \begin{equation}
    \label{eq:zero-lightlike}
    \boxed{[H ,\B] = \B + \P \qquad [H ,\P] = \P \qquad [H ,Z] = 2Z.}
  \end{equation}
  This Lie algebra $\hat\g$ is a non-central extension
  \begin{equation}
    \begin{CD}
      0 @>>> \RR\left<Z\right> @>>> \hat\g @>>> \hat\h @>>> 0,
    \end{CD}
  \end{equation}
  of the Lie algebra $\hat\h$ given by
  Equation~\eqref{eq:3-2-branch}.

\item $u_2^2 - u_2 u_4 < 0$:
  \begin{equation}
    \label{eq:zero-timelike}
    \boxed{[H , \B] = \alpha \B + \P \qquad [H ,\P] = \alpha \P - \B
      \qquad [H ,Z] = 2\alpha Z,}
    \end{equation}
    for $\alpha > 0$.  This Lie algebra $\hat\g$ is a non-central
    extension
    \begin{equation}
      \begin{CD}
        0 @>>> \RR\left<Z\right> @>>> \hat\g @>>> \hat\h @>>> 0,
      \end{CD}
    \end{equation}
    of the Lie algebra $\hat\h$ given by
    Equation~\eqref{eq:2-2-branch}.  The limiting case $\alpha = 0$ is the central
  extension of the euclidean Newton algebra discussed above
  (corresponding to $u_1 = 0$).
\end{enumerate}

\subsection{Spacelike branch deformations}
\label{sec:spac-branch-deform-1}

Here $u_5 =1$ and $u_6=u_7=0$.  The system~\eqref{eq:quadrics-cext} of
quadrics becomes $u_1 = u_3 = u_4 = 0$, so the deformation becomes
\begin{equation}
  \varphi_1 = - u_2 (\ct_{14} - \ct_{16}) + \ct_3 + \ct_{20}- \ct_{22},
\end{equation}
leading to the Lie brackets
\begin{equation}
  \begin{aligned}[m]
    [H , \B] &= u_2 \B\\
    [H ,\P] &= -u_2 \P\\
  \end{aligned}
  \qquad\qquad
  \begin{aligned}[m]
    [Z,\B] &= -\B\\
    [Z,\P] &= \P\\
  \end{aligned}
  \qquad\qquad
  [\B,\P] = Z + \R.
\end{equation}
It follows that $H  + u_2 Z$ is central, so that this deformation is
a trivial central extension of the Lie algebra
\begin{equation}\label{eq:triv-so41}
  \boxed{[Z,\B] = -\B \qquad [Z,\P] = \P \qquad\text{and}\qquad  [\B,\P] = Z + \R,}
\end{equation}
which is isomorphic (with $Z$ here playing the role of $-H $ there) to the
Lie algebra in \eqref{eq:1-1-branch}; that is, to $\so(4,1)$.

\subsection{Timelike branches deformations}
\label{sec:timel-branch-deform-1}

Let us introduce $\varepsilon= \pm 1$ and treat both branches
simultaneously.  Here $u_5 = 0$ and $u_6 = u_7 = \varepsilon$.  The
system~\eqref{eq:quadrics-cext} of quadrics says that $u_1 = u_2 = 0$
and that $u_3 = u_4$, so that the deformation is given by
\begin{equation}
  \varphi_1 = u_3 \left(\ct_{13} - \ct_{17}\right) + \varepsilon (
  \ct_6 + \ct_9 - \ct_{19} + \ct_{23}),
\end{equation}
with Lie brackets
\begin{equation}
  \begin{aligned}[m]
    [H ,\B] &= - u_3 \P\\
    [H ,\P] &=  u_3 \B\\
  \end{aligned}
  \qquad\qquad
  \begin{aligned}[m]
    [Z,\B] &= \varepsilon\P\\
    [Z,\P] &= -\varepsilon \B\\
  \end{aligned}
  \qquad\qquad
  \begin{aligned}[m]
    [\B,\B] &= \varepsilon\R\\
    [\P,\P] &= \varepsilon\R.\\
  \end{aligned}
\end{equation}
It follows that $\varepsilon H  + u_3 Z$ is central, and we have a
trivial central extension to the Lie algebra with brackets
\begin{equation}\label{eq:triv-so}
\boxed{  \begin{aligned}[m]
    [Z,\B] &= \varepsilon\P\\
    [Z,\P] &= -\varepsilon \B\\
  \end{aligned}
  \qquad\qquad
  \begin{aligned}[m]
    [\B,\B] &= \varepsilon\R\\
    [\P,\P] &= \varepsilon\R.\\
  \end{aligned}
  \qquad\qquad
  [\B,\P] = Z,}
\end{equation}
which (again $Z$ playing the role of $\varepsilon H $) is isomorphic
to the Lie algebra in \eqref{eq:2-1-branch}; that is, to $\so(5)$
or $\so(3,2)$ depending on $\varepsilon$.

\subsection{Lightlike branches deformations}
\label{sec:lightl-branch-deform-1}

We again introduce $\varepsilon = \pm 1$ and treat both branches
simultaneously.  We have that $u_5 = u_6 = 0$ and $u_7 =
\varepsilon$.  The system~\eqref{eq:quadrics-cext} of quadrics imply
that $u_3=0$ and $u_2 = - u_1$, so that the deformation ends up being
\begin{equation}
  \varphi_1 = 2 u_1 (\ct_{14} + \ct_{24}) - u_4 \ct_{17} + \varepsilon
  (\ct_6 + \ct_{23}),
\end{equation}
with Lie brackets
\begin{equation}
  \begin{aligned}[m]
    [H ,\B] &= - u_4 \P\\
    [H ,\P] &= 2 u_1 \P\\
  \end{aligned}
  \qquad\qquad
  \begin{aligned}[m]
    [H ,Z] &= 2 u_1 Z\\
    [Z,\B] &= \varepsilon \P\\
  \end{aligned}
  \qquad\qquad
  [\B,\B] = \varepsilon \R.
\end{equation}
Let us change basis from $(H ,Z)$ to $(H  + \varepsilon u_4 Z, Z)$.
In this new basis, the non-zero brackets are
\begin{equation}
  \begin{aligned}[m]
    [H ,\P] &= 2 u_1 \P\\
    [H ,Z] &= 2 u_1 Z\\
  \end{aligned}
  \qquad\qquad
  \begin{aligned}[m]
    [Z,\B] &= \varepsilon \P\\
    [\B,\B] &= \varepsilon \R.
  \end{aligned}
\end{equation}

We must distinguish two cases, depending on whether or not $u_1 = 0$.

\subsubsection{$u_1=0$ subbranch}
\label{sec:u_1=0-branch}

If $u_1 =0$, then (the new) $H $ is central and we obtain a trivial
central extension of the Lie algebra with brackets
\begin{equation}
  \label{eq:trivial-iso}
  \boxed{[\B,\P] = Z \qquad [Z,\B] = \varepsilon \P
    \qquad\text{and}\qquad [\B,\B] = \varepsilon \R,}
\end{equation}
which is isomorphic to either the euclidean or Poincaré Lie algebras
(with $Z$ playing the role of $-\varepsilon H $) depending on
$\varepsilon$.

\subsubsection{$u_1 \neq 0$ subbranch}
\label{sec:u_1-neq-0-1}

If $u_1 \neq 0$, then we may rescale $H $ to set $u_1 = 1$ and arrive
at the Lie algebra with non-zero brackets
\begin{equation}
  \label{eq:conformal}
  \boxed{
    \begin{aligned}[m]
    [H ,\P] &= \P\\
    [H ,Z] &= Z\\
  \end{aligned}
  \qquad\qquad
  \begin{aligned}[m]
    [\B,\P] &= Z\\
    [Z,\B] &= \varepsilon \P\\
  \end{aligned}
  \qquad\qquad
  [\B,\B] = \varepsilon \R,}
\end{equation}
which leads to a deformation isomorphic to either the conformal
euclidean or conformal Poincaré Lie algebras, depending on the sign of
$\varepsilon$.  In other words, $\co(4) \ltimes \RR^4$ or $\co(3,1)
\ltimes \RR^{3,1}$, with $Z$ playing the role of the fourth
translation and $H $ playing the role of the dilatation.

\subsection{Invariant inner products}
\label{sec:invar-inner-prod-1}

We shall now analyse the existence of invariant inner products on the
Lie algebras determined in this section, as we did in
Section~\ref{sec:invar-inner-prod} for the kinematical Lie algebras
classified in Section~\ref{sec:deform-stat-kinem}.  In some cases we
will appeal to a general result about associative inner products on
Lie algebras, which says that the center $Z(\g)$ of a Lie algebra with
an invariant inner product is the perpendicular of the first derived
ideal $\g' = [\g,\g]$; that is $\g' = Z(\g)^\perp$.  Therefore if $\g$
is such that $Z(\g) =0$ and $\g' \subsetneq \g$, then $\g$ cannot
admit an invariant inner product.  This is precisely the situation of
the Lie algebras in the bottom third (below the line) of
Table~\ref{tab:defs-ce-kla-summary}.

The first Lie algebra in the table (with brackets given by
\eqref{eq:static-centrext}) does not admit an invariant inner
product.  Indeed, if $(-,-)$ is an associative symmetric bilinear
form, it follows that
\begin{equation}
  \delta_{ij} (Z,Z) = ([B_i,P_j],Z) = (B_i, [P_j,Z]) = 0
\end{equation}
and
\begin{equation}
  \delta_{ij} (Z,H) = ([B_i,P_j],H) = (B_i, [P_j,H]) = 0,
\end{equation}
so that $(Z,-) = 0$.  The exact same calculation shows that in the
Bargmann algebra \eqref{eq:bargmann} any associative symmetric
bilinear form has $(Z,-) = 0$.  A very similar argument shows that the
trivial central extensions of the euclidean
and Poincaré algebras \eqref{eq:trivial-iso} do not admit invariant
inner products either.  Indeed, if $(-,-)$ is any associative
symmetric bilinear form, then
\begin{equation}
  \delta_{ij} (H,H) = ([B_i,P_j],H) = (B_i, [P_j,H]) = 0
\end{equation}
and
\begin{equation}
  \delta_{ij} (H,Z) = ([B_i,P_j],Z) = (B_i, [P_j,Z]) = 0,
\end{equation}
so that $(H,-) = 0$.
The trivial central extensions of $\so(4,1)$, $\so(5)$ and $\so(3,2)$
do admit invariant inner products by taking the Killing form on the
simple factor and some non-zero value for $(Z,Z)$.

Finally, we treat the centrally extended Newton algebras.  The two
cases are very similar, so we give details only for the case of the
lorentzian algebra \eqref{eq:celornewton}.  Let $(-,-)$ be an
associative symmetric bilinear form.  We will show that $(B_i,-) = 0$,
so that it is degenerate.  First of all, by rotational invariance,
$(B_i, H) = (B_i, Z) = 0$.  Let us calculate the others:
\begin{equation}
  \begin{split}
    (B_i, R_j) &= ([H,B_i], R_j) = - ([B_i, H], R_j) = - (B_i, [H,R_j]) = 0\\
    (B_i, B_j) &= ([H,B_i], B_j) = (H, [B_i,B_j]) = 0\\
    \epsilon_{ij\ell} (B_\ell,P_k) &= ([R_i,B_j],P_k) = (R_i, [B_j, P_k]) =
    \delta_{jk} (R_i, Z) = 0.
  \end{split}
\end{equation}
The euclidean case \eqref{eq:ceeucnewton} is similar.  In summary,
only the trivial central extensions of the simple kinematical Lie
algebras $\so(4,1)$, $\so(5)$ and $\so(3,2)$ admit invariant inner
products.

\subsection{Summary}
\label{sec:summary-1}

The results of this section are partially known and partially new.
They extend and in at least one case correct the results of our 1989
paper \cite{JMFGalilean} on the deformations of the galilean and
Bargmann algebras.  Table~\ref{tab:defs-ce-kla-summary} lists our
results.  All of these Lie algebras share the following Lie brackets
(in abbreviated notation):
\begin{equation}
  [\R,\R] = \R \qquad [\R,\B] = \B \qquad [\R, \P] = \P \qquad [\R,
  H ] = 0  \qquad\text{and}\qquad [\R, Z] =0.
\end{equation}
In the table we will only list any additional non-zero brackets.  In
some cases we have interchanged $Z$ and $H $ to make the notation
more uniform.  The table is divided into three: the top third
consists of (non-trivial) central extensions, the middle third of
trivial central extensions  and the bottom third of non-central
extensions of kinematical Lie algebras.

\begin{table}[h!]\tiny
  \centering
  \caption{Deformations of the centrally extended static kinematical Lie algebra}
  \label{tab:defs-ce-kla-summary}
  \begin{tabular}{l|*{5}{>{$}l<{$}}|l|c}
    \multicolumn{1}{c|}{Eq.} & \multicolumn{5}{c|}{Nonzero Lie brackets} & \multicolumn{1}{c|}{Comments} & \multicolumn{1}{c}{Metric?}\\\hline
    \ref{eq:static-centrext} & [\B,\P] = Z & & & & & centrally extended static  & \\
    \ref{eq:celornewton} & [\B,\P] = Z & [H , \B] = \B & [H ,\P] = -\P & & & central extension of lorentzian Newton  & \\
    \ref{eq:ceeucnewton} & [\B,\P] = Z & [H , \B] = \P & [H ,\P] = -\B & & & central extension of euclidean Newton  & \\
    \ref{eq:bargmann} & [\B,\P] = Z & [H , \B] = - \P & & & & Bargmann  & \\\hline
    \ref{eq:trivial-iso} & [\B,\P] = H  & [H , \B] = \P & & & [\B,\B] = \R & $\e\oplus\RR Z$ & \\
    \ref{eq:trivial-iso} & [\B,\P] = H  & [H , \B] = - \P & & & [\B,\B] = - \R & $\p\oplus\RR Z$ & \\
    \ref{eq:triv-so41} & [\B,\P] = H  + \R & [H , \B] = -\B & [H , \P] = \P & & & $\so(4,1) \oplus \RR Z$ & \checkmark\\
    \ref{eq:triv-so} & [\B,\P] = H  & [H ,\B] = \P & [H , \P] = - \B &  [\P, \P] = \R & [\B,\B] = \R & $\so(5)\oplus \RR Z$ & \checkmark\\
    \ref{eq:triv-so} & [\B,\P] = H  & [H ,\B] = - \P & [H , \P] = \B &  [\P, \P] = - \R & [\B,\B] = - \R & $\so(3,2)\oplus \RR Z$ & \checkmark\\\hline
    \ref{eq:zero-spacelike} & [\B,\P] = Z & [H , \B] = \B & [H ,\P] = \P & [H ,Z] = 2 Z& &  & \\
    \ref{eq:zero-spacelike} & [\B,\P] = Z & [H , \B] = \gamma\B & [H ,\P] = \P & [H ,Z] = (\gamma+1) Z& & $\gamma \in (-1,1)$  & \\
    \ref{eq:zero-lightlike} & [\B,\P] = Z & [H , \B] = \B + \P & [H , \P] = \P & [H ,Z] = 2 Z & &  & \\
    \ref{eq:zero-timelike} & [\B,\P] = Z & [H , \B] = \alpha \B + \P & [H ,\P] = -\B + \alpha \P & [H ,Z] = 2\alpha Z& & $\alpha > 0$  & \\
    \ref{eq:conformal} & [\B,\P] = Z & [Z,\B] = \P & [H ,\P] = \P & [H , Z] = Z & [\B,\B] = \R & $\co(4)\ltimes \RR^4$ & \\
    \ref{eq:conformal} & [\B,\P] = Z & [Z,\B] = - \P & [H ,\P] = \P & [H , Z] = Z & [\B,\B] = - \R & $\co(3,1) \ltimes \RR^{3,1}$ & \\
  \end{tabular}
\end{table}

\section{Conclusions}
\label{sec:conclusions}

We have presented a deformation theory approach to the classification
of kinematical Lie algebras (in $3+1$ dimensions) as deformations of
the static kinematical Lie algebra: the one where all brackets except
those which define it as a kinematical Lie algebra are zero.  We saw
that all deformations of the static Lie algebra are necessarily
kinematical.  This recovers the classical result of Bacry and Nuyts
\cite{MR857383}.  The static kinematical Lie algebra admits a
one-dimensional central extension and we also determine all
deformations of that algebra.  In the process we recover some known
Lie algebras -- namely, those which are (trivial or non-trivial)
central extensions of kinematical Lie algebras -- but also some Lie
algebras which are \emph{non-central} extensions of kinematical Lie
algebras.  This should not come as a surprise, since deformation and
central extension do not commute, hence there is no reason to expect
that deforming the central extension of a Lie algebra $\g$ one should
recover the central extension of a deformation of $\g$.

The results are summarised in two tables: Table~\ref{tab:kla-summary}
contains the kinematical Lie algebras and is to be compared with
Table~1 in \cite{MR857383}, whereas Table~\ref{tab:defs-ce-kla-summary}
contains the deformations of the centrally extended static kinematical
Lie algebras.  The notation employed in these tables is an abbreviated
notation borrowed from \cite{MR857383}.

This paper lays the groundwork for two companion papers: one
\cite{JMFKinematicalHD} where we obtain the analogous classifications
as in this paper but in dimension $D+1$ for all $D\geq 4$, and another
\cite{TAJMFKinematical2D} where we classify kinematical Lie algebras
in dimension $2+1$.  The three papers have been separated because they
differ substantially in the technicalities, despite sharing a similar
methodology.  This series of papers lay the foundations to the
classification of homogeneous spacetimes of kinematical Lie
algebras in all dimensions, which is work in progress in collaboration
with Stefan Prohazka.

It should be mentioned that there exists a classification of
kinematical Lie superalgebras \cite{CampoamorStursberg:2008hm} in
$3+1$ dimensions, which would be interesting to extend to other
dimensions.

\section*{Acknowledgments}
\label{sec:acknowledgments}

This research is partially supported by the grant ST/L000458/1
``Particle Theory at the Higgs Centre'' from the UK Science and
Technology Facilities Council.  I'm grateful to the referee for a
careful reading of the manuscript and spotting a number of
typographical errors in formulae.

\appendix

\section{Lie algebra cohomology}
\label{sec:lie-algebra-cohom}

In this appendix we review very briefly the definition of Lie algebra
cohomology as introduced by Chevalley and Eilenberg in
\cite{ChevalleyEilenberg}.

\subsection{Chevalley--Eilenberg complex}
\label{sec:chev-eilenb-compl}

The cohomology of the Lie algebra of a Lie group $G$ can be calculated
using the Chevalley--Eilenberg complex, which is isomorphic to the
subcomplex of the de~Rham complex of $G$ consisting of left-invariant
differential forms.  There is also a purely algebraic description
which takes as starting data a Lie algebra and a representation.

Let $\g$ be a (finite-dimensional, real) Lie algebra and $\m$ a
module.  If $X \in \g$ and $v \in \m$, we will let $X v \in \m$ denote
the action of $X$ on $v$.  Being a module, it satisfies
$X (Y v) - Y (X v) = [X,Y] v$, for all $X,Y\in\g$ and $v\in\m$.  The
cochains in the Chevalley--Eilenberg complex are skew-symmetric
multilinear maps $\Lambda^p \g \to \m$ where $p$ runs from $0$ to
$\dim\g$.  Let $C^p(\g;\m) = \Lambda^p\g^*\otimes\m$ denote the space
of $p$-cochains.  The differential
$\d: C^p(\g;\m) \to C^{p+1}(\g;\m)$ is determined by its action
on $\g^*$ and $\m$ and extending it as an odd derivation over the
wedge product.  If $v \in \m$, then $\d v \in \g^*\otimes\m$ is
given by
\begin{equation}
  \d v(X) = X v
\end{equation}
and if $\alpha \in \g^*$, $\d\alpha \in \Lambda^2\g^*$ is given by
\begin{equation}
  \d\alpha(X,Y) = -\alpha([X,Y]),
\end{equation}
for all $X,Y \in \g$.  Since $\d$ is an odd derivation, $\d^2 =
\tfrac12 [\d,\d]$ is an even derivation, so it is also determined by
its action on generators.  On $v \in \m$, $\d^2m = 0$ using that $\m$
is a $\g$-module, whereas on $\alpha \in \g^*$, $\d^2\alpha = 0$ by
virtue of the Jacobi identity of $\g$.  Therefore $\d^2=0$.

Let $X_i$ be a basis for $\g$ and $\alpha^i$ the canonically dual
basis for $\g^*$.  Let $[X_i,X_j]=f_{ij}{}^k X_k$ define the structure
constants of $\g$ relative to this choice of basis.  Then we can write
the differentials above as follows:
\begin{equation}
  \d v = \alpha^i \otimes X_i v \qquad\text{and}\qquad \d\alpha^k = -
  \tfrac12 f_{ij}{}^k \alpha^i \wedge \alpha^j
\end{equation}
and we extend it by
\begin{equation}
  \d(\alpha \wedge \beta \otimes v) = \d\alpha \wedge \beta \otimes v +
  (-1)^{|\alpha|} \alpha \wedge \d\beta \otimes v +
  (-1)^{|\alpha|+|\beta|} \alpha \wedge \beta \wedge \d v,
\end{equation}
for all homogeneous $\alpha,\beta \in \Lambda^\bullet\g^*$ and $v \in \m$.

The relevant complex when computing Lie algebra deformations of $\g$
is $C^\bullet(\g;\g)$ where $\m = \g$ is the adjoint representation.
In this case the first three differentials $\d: C^p(\g;\g) \to
C^{p+1}(\g;\g)$ for $p=0,1,2$ are given explicitly, for $X,Y,Z \in \g$,
$\beta \in C^1(\g;\g)$ and $\mu \in C^2(\g;\g)$, by
\begin{equation}
  \begin{split}
    \d X(Y) &= - [X,Y]\\
    \d \beta(X,Y) &= [X,\beta(Y)] - [Y,\beta(X)] - \beta([X,Y])\\
    \d \mu(X,Y,Z) &= [X,\mu(Y,Z)] - \mu([X,Y],Z) + \text{cyclic}.
  \end{split}
\end{equation}

In this paper, however, we are interested not in all Lie algebra
deformations, but only in deformations within the class of kinematical
Lie algebras.  The complex $C^\bullet(\g;\g)$ seems too big at face
value and we should work instead with a \textbf{relative} subcomplex.
Let $\s$ be a Lie subalgebra of $\g$ (in the case which interests us
in this paper, $\s \cong \so(3)$ is the rotational subalgebra).  We
limit ourselves to deformations where the brackets involving $\s$ are
not modified.  This means that if $\varphi \in C^2(\g;\g)$ is the
deformation, we require that $\iota_X \varphi = 0$ for all $X \in \s$
and we also require that $\varphi$ be $\s$-invariant, which follows
from the Jacobi identity involving one element from $\s$.  These two
conditions are equivalent to $\iota_X \alpha = 0$ and
$\iota_X \d\alpha = 0$ for all $X \in \s$, which defines the relative
subcomplex $C^\bullet(\g,\s;\g)$.  For the static kinematical Lie
algebra $\g$ (and also for its universal central extension), the
rotational subalgebra $\s$ has a complementary ideal $\h$ and then the
relative subcomplex $C^\bullet(\g,\s;\g)$ is isomorphic to the
subcomplex $C^\bullet(\h;\g)^\s$ consisting of the $\s$-invariant
elements of the Chevalley--Eilenberg complex of the Lie algebra $\h$
relative to the representation $\g$.  As we will now briefly recall, a
celebrated theorem of Hochschild and Serre says that there is a close
relation between the cohomology of $C^\bullet(\h;\g)^\s$ and of
$C^\bullet(\g;\g)$.  In particular, for the static kinematical Lie
algebra (and also for its universal central extension), every
deformation will  necessarily be kinematical.

\subsection{The Hochschild--Serre spectral sequence}
\label{sec:hochschild-serre}

In \cite{MR0054581} Hochschild and Serre proved a factorisation
theorem that in many cases simplifies the calculation of Lie algebra
cohomology groups. Let $\g$ be a finite-dimensional real Lie algebra
and $\h$ an ideal such that the quotient Lie algebra $\s = \g/\h$ is
semisimple. Let $\m$ denote a $\g$-module, which is then also an
$\h$-module.  Hochschild and Serre use the ideal $\h$ to define a
filtration of the cochains $C^\bullet(\g;\m)$, whose associated
spectral sequence degenerates at the second page yielding the
following isomorphism:
\begin{equation}
  H^n(\g;\m) \isom \bigoplus_{i=0}^{n} H^{n-i}(\s;\RR)\tensor
  H^{i}(\h;\m)^{\s},
\end{equation}
where the superscript $\s$ denotes $\s$-invariants.  Since $\s$
is semisimple, it acts reducibly on the cochains $C^\bullet(\h;\m)$ and hence
the $\s$-invariant cohomology can be computed from the $\s$-invariant
cochains.

Moreover, from the Whitehead lemmas (see, e.g.,
\cite[§III.10]{MR559927}), $H^1(\s;\RR) = H^2(\s;\RR)= 0$.  If, in
addition, $\s$ is simple then $H^3(\s;\RR)\isom \RR$.  Hence for $\s$
simple, the first few $H^\bullet(\g;\g)$ are as follows
\begin{equation}
  \begin{aligned}
    H^0(\g;\g) &\isom Z(\g)\\
    H^1(\g;\g) &\isom H^1(\h;\g)^{\s}\\
  \end{aligned}
  \qquad\qquad
  \begin{aligned}
    H^2(\g;\g) &\isom H^2(\h;\g)^{\s}\\
    H^3(\g;\g) &\isom H^3(\h;\g)^{\s}\oplus Z(\g),\\
  \end{aligned}
\end{equation}
where $Z(\g)$ denotes the center of $\g$.  In particular, the
infinitesimal deformations of $\g$ are such that the brackets involving
$\s$ are not modified.  This, of course, is a consequence of the
well-known rigidity of semisimple Lie algebras and their
finite-dimensional modules.

\subsection{Central extension of the static kinematical Lie algebra}
\label{sec:centr-extens-stat}

As an application of the Hochschild--Serre factorisation theorem, let
us calculate the universal central extension of the static kinematical
Lie algebra.

The static kinematical Lie algebra $\g$ is a ten-dimensional Lie
algebra with generators $R_i$, $B_i$, $P_i$ and $H $, with the
following non-zero Lie brackets:
\begin{equation}
  [R_i, R_j] = \epsilon_{ijk} R_k \qquad
  [R_i, B_j] = \epsilon_{ijk} B_k \qquad\text{and}\qquad
  [R_i, P_j] = \epsilon_{ijk} P_k.
\end{equation}
In other words, $\g$ is isomorphic to the semidirect product of the
simple Lie algebra $\so(3)$ (spanned by the $R_i$) and an abelian Lie
algebra transforming as the representation $2 V \oplus \RR$, where $V$
is the $3$-dimensional vector representation and $\RR$ is the trivial
representation.  It is often convenient to abbreviate the Lie bracket
as follows:
\begin{equation}
  [\R, \R] = \R \qquad   [\R, \B] = \B \qquad\text{and}\qquad
  [\R,\P] = \P,
\end{equation}
which does not lead to any ambiguity as there is (up to scale) only
one $\so(3)$-equivariant map $V \otimes V \to V$.

Central extensions of $\g$ are classified by the second
Chevalley--Eilenberg cohomology group $H^2(\g;\RR)$, which by
Hochschild--Serre is isomorphic to $H^2(\h;\RR)^{\s}$, where $\h$ is
the abelian ideal generated by $B_i, P_i, H $ and $\s \cong \so(3)$ is
the simple subalgebra generated by $R_i$.  Since $\s$ is simple, and
hence reductive, we may calculate the $\s$-invariant cohomology from
the $\s$-invariant subcomplex: $C^p(\h;\RR)^{\s}=
\Hom_\s(\Lambda^p\h,\RR)$.  By inspection, $C^p(\h;\RR)^{\s}$, for
$p=1,2$, are one-dimensional with basis $\eta $ and $\pi^i \wedge
\beta^i$, respectively, where $\beta^i,\pi^i,\eta $ are the basis for
$\h^*$ canonically dual to the basis $B_i, P_i, H $ for $\h$.  Since
$\h$ is abelian, the Chevalley--Eilenberg differential is identically
zero and hence
\begin{equation}
  H^2(\h;\RR)^{\s} \cong \RR\left<[\pi^i \wedge \beta^i]\right>,
\end{equation}
so that there is a one-dimensional universal central extension with
Lie bracket
\begin{equation}
  \label{eq:static-CE}
  [B_i, P_j] = \delta_{ij} Z \qquad\text{(abbreviated as $[\B,\P]=Z$)}
\end{equation}
where $Z$ is the central generator.

\section{Cochains for the static kinematical Lie algebra}
\label{sec:coch-stat-kinem}

In this appendix we list the relevant cochains in the complex
calculating $H^2(\g;\g)$ for the static kinematical Lie algebra $\g$
with basis $R_i, B_i, P_i, H $.  The ideal $\h$ is spanned by
$B_i,P_i,H $ with simple quotient $\s$, isomorphic to the subalgebra
generated by $R_i$.  The canonical dual basis for $\h^*$ is given by
$\beta^i,\pi^i,\eta $.  By Hochschild--Serre, it suffices to calculate
the cohomology of the $\s$-invariant complex $C^\bullet(\h;\g)^{\s}$.
The relevant cochains are tabulated below using an abbreviated
notation where we have omitted $\otimes$, $\wedge$ and any indices.
For example, $\beta R = \beta^i \otimes R_i$, $\tfrac12 \beta\beta R =
\tfrac12 \epsilon_{ijk} \beta^i \wedge \beta^j \otimes R_k$ and
$\beta\pi\pi B = \beta^i \wedge \pi^i \wedge \pi^j \otimes B_j$.

\begin{table}[h!]
  \centering
  \caption{Basis for $C^1(\h;\g)^{\s}$}
  \label{tab:basis-C1}
  \begin{tabular}{*{7}{>{$}c<{$}}}
    a_1& a_2 & a_3 & a_4 & a_5 &a_6 & a_7 \\\hline
    \beta R & \beta B & \beta P & \pi R & \pi B & \pi P & \eta  H 
  \end{tabular}
\end{table}

\begin{table}[h!]
  \centering
  \caption{Basis for $C^2(\h;\g)^{\s}$}
  \label{tab:basis-C2}
  \begin{tabular}{*{8}{>{$}c<{$}}}
    c_1 & c_2 & c_3 & c_4 & c_5 & c_6 & c_7 & c_8 \\\hline
    \beta\pi H  & \eta  \beta R & \eta  \beta B & \eta  \beta P & \eta  \pi R & \eta  \pi B & \eta  \pi P & \tfrac12 \beta\beta R\\[10pt]
    c_9 & c_{10} & c_{11} & c_{12} & c_{13} & c_{14} & c_{15} & c_{16} \\\hline
    \tfrac12 \beta\beta B & \tfrac12 \beta\beta P & \beta \pi R & \beta \pi B & \beta \pi P & \tfrac12 \pi \pi R & \tfrac12 \pi \pi B & \tfrac12 \pi \pi P
  \end{tabular}
\end{table}

\begin{table}[h!]
  \centering
  \caption{Basis for $C^3(\h;\g)^{\s}$}
  \label{tab:basis-C3}
  \begin{tabular}{*{10}{>{$}c<{$}}}
    b_1 & b_2 & b_3 & b_4 & b_5 & b_6 & b_7 & b_8 & b_9 & b_{10} \\\hline
    \beta\pi\beta P & \eta \beta\beta B & \eta \beta\pi P & \eta \beta\pi B & \eta \pi\pi P & \eta \beta\beta P & \eta \beta\beta R & \beta\pi\pi P& \beta\pi\beta B& \eta  \beta\pi R\\[10pt]
     b_{11} &  b_{12} &  b_{13} &  b_{14} &  b_{15} &  b_{16} &  b_{17} &  b_{18} &  b_{19}&  b_{20} \\\hline
    \beta\pi\pi B & \beta\beta\beta H  & \eta \beta\pi H  & \eta \pi\pi B & \beta\pi\beta R & \beta\pi\pi R & \eta \pi\pi R & \beta\beta\pi H  & \beta\pi\pi H  & \pi\pi\pi H 
  \end{tabular}
\end{table}

The Chevalley--Eilenberg differential is defined on generators in such a way that it is zero except for
\begin{equation}
  \d R_i = -\epsilon_{ijk} (\beta^j B_k + \pi^j P_k),
\end{equation}
from where we can calculate the differential on cochains. Using the notation in the above tables of cochains, the non-zero differentials are:
\begin{equation}
  \label{eq:CE-static}
  \begin{aligned}[m]
    \d a_1 &= 2 c_9 + c_{13}\\
    \d a_4 &= c_{12} + 2 c_{16}
  \end{aligned}
  \qquad\qquad
  \begin{aligned}[m]
    \d c_2 &= -b_2 - b_3\\
    \d c_5 &= -b_4 - b_5
  \end{aligned}
  \qquad\qquad
  \begin{aligned}[m]
    \d c_8 &= b_1 \\
    \d c_{11} &= b_8 - b_9
  \end{aligned}
  \qquad\qquad
  \begin{aligned}[m]
    \d c_{14} &= -b_{11}.\\
  \end{aligned}
\end{equation}

Finally, we work out (the restriction of) the Nijenhuis--Richardson bracket
\begin{equation}
  [\![-,-]\!] : C^2(\h;\g)^{\s} \times C^2(\h;\g)^{\s} \to C^3(\h;\g)^{\s}
\end{equation}
on the above basis of cochains:
\begin{equation}
  \label{eq:NR-static-explicit}
  \small
  \begin{aligned}[m]
    [\![c_1,c_2]\!] &= b_{15}\\
    [\![c_1,c_3]\!] &= b_9 + b_{13}\\
    [\![c_1,c_4]\!] &= b_1\\
    [\![c_1,c_5]\!] &= b_{16}\\
    [\![c_1,c_6]\!] &= b_{11}\\
    [\![c_1,c_7]\!] &= b_8 + b_{13}\\
    [\![c_1,c_9]\!] &= \tfrac12 b_{18}\\
    [\![c_1,c_{10}]\!] &= -\tfrac12 b_{12}\\
    [\![c_1,c_{12}]\!] &= b_{19}\\
    [\![c_1,c_{13}]\!] &= b_{18}\\
    [\![c_1,c_{15}]\!] &= b_{20}\\
    [\![c_1,c_{16}]\!] &= -\tfrac12 b_{20}\\
    [\![c_2,c_9]\!] &= \tfrac12 b_7\\
    [\![c_2,c_{12}]\!] &= - b_{10}\\
    [\![c_2,c_{15}]\!] &= - \tfrac12 b_{17}\\
  \end{aligned}
  \qquad
  \begin{aligned}[m]
    [\![c_3,c_8]\!] &= b_7\\
    [\![c_3,c_9]\!] &= \tfrac32 b_2\\
    [\![c_3,c_{10}]\!] &= b_6\\
    [\![c_3,c_{11}]\!] &= b_{10}\\
    [\![c_3,c_{13}]\!] &= b_3\\
    [\![c_3,c_{15}]\!] &= -\tfrac12 b_{14}\\
    [\![c_4,c_9]\!] &= \tfrac12 b_6\\
    [\![c_4,c_{11}]\!] &= b_7\\
    [\![c_4,c_{12}]\!] &= b_2 - b_3\\
    [\![c_4,c_{13}]\!] &= b_6 \\
    [\![c_4,c_{14}]\!] &= b_{10} \\
    [\![c_4,c_{15}]\!] &= b_4 - \tfrac12 b_5\\
    [\![c_4,c_{16}]\!] &= b_3\\
    [\![c_5,c_{10}]\!] &= -\tfrac12 b_7 \\
    [\![c_5,c_{13}]\!] &= - b_{10}\\
    [\![c_5,c_{16}]\!] &= - \tfrac12 b_{17} \\
  \end{aligned}
  \qquad
  \begin{aligned}[m]
    [\![c_6, c_8]\!] &= b_{10} \\
    [\![c_6, c_9]\!] &= b_4 \\
    [\![c_6, c_{10}]\!] &= b_3 - \tfrac12 b_2\\
    [\![c_6, c_{11}]\!] &= b_{17} \\
    [\![c_6, c_{12}]\!] &= b_{14} \\
    [\![c_6, c_{13}]\!] &= b_5 - b_4 \\
    [\![c_6, c_{16}]\!] &= -\tfrac12 b_{14}\\
    [\![c_7, c_{10}]\!] &= -\tfrac12 b_6\\
    [\![c_7, c_{11}]\!] &= b_{10} \\
    [\![c_7, c_{12}]\!] &= b_4 \\
    [\![c_7, c_{14}]\!] &= b_{17} \\
    [\![c_7, c_{15}]\!] &= b_{14} \\
    [\![c_7, c_{16}]\!] &= \tfrac12 b_5\\
    [\![c_8, c_{12}]\!] &= b_{15} \\
    [\![c_8, c_{15}]\!] &= b_{16} \\
  \end{aligned}
  \qquad
  \begin{aligned}[m]
    [\![c_9, c_{11}]\!] &= - b_{15}\\
    [\![c_9, c_{13}]\!] &= - b_1 \\
    [\![c_9, c_{15}]\!] &= b_{11}\\
    [\![c_{10}, c_{12}]\!] &= b_1\\
    [\![c_{10}, c_{14}]\!] &= - b_{15}\\
    [\![c_{10}, c_{15}]\!] &= b_8 - b_9\\
    [\![c_{10}, c_{16}]\!] &= - b_1\\
    [\![c_{11}, c_{12}]\!] &= -b_{16}\\
    [\![c_{11}, c_{13}]\!] &= b_{15}\\
    [\![c_{11}, c_{16}]\!] &= b_{16}\\
    [\![c_{12}, c_{12}]\!] &= -2 b_{11}\\
    [\![c_{12}, c_{13}]\!] &= -b_8 + b_9\\
    [\![c_{12}, c_{16}]\!] &= b_{11}\\
    [\![c_{13}, c_{13}]\!] &= 2 b_1\\
    [\![c_{13}, c_{14}]\!] &= -b_{16}\\
    [\![c_{13}, c_{15}]\!] &= -b_{11}\\
  \end{aligned}
\end{equation}

\section{Cochains for the centrally extended static kinematical Lie algebra}
\label{sec:coch-centr-stat-kinem}

In this appendix we list the relevant cochains in the complex
calculating $H^2(\tg;\tg)$ for the centrally extended static
kinematical Lie algebra $\tg$ with basis $R_i, B_i, P_i, H , Z$.  The
ideal $\th$ is spanned by $B_i,P_i,H , Z$ with simple quotient $\s$,
isomorphic to the subalgebra generated by $R_i$.  The canonical dual
basis for $\th^*$ is given by $\beta^i,\pi^i,\eta ,\zeta$.  By
Hochschild--Serre, it suffices to calculate the cohomology of the
$\s$-invariant complex $C^\bullet(\th;\tg)^{\s}$.  The relevant
cochains are tabulated below using the same abbreviated notation as in
the previous appendix.

\begin{table}[h!]
  \centering
  \caption{Basis for $C^1(\th;\tg)^{\s}$}
  \label{tab:basis-C1-ext}
  \begin{tabular}{*{10}{>{$}c<{$}}}
    \at_1& \at_2 & \at_3 & \at_4 & \at_5 &\at_6 & \at_7 & \at_8 & \at_9 & \at_{10}\\\hline
    \zeta H  & \zeta Z & \eta  H  & \eta  Z & \beta R & \beta B & \beta P & \pi R & \pi B & \pi P
  \end{tabular}
\end{table}

\begin{table}[h!]
  \centering
  \caption{Basis for $C^2(\th;\tg)^{\s}$}
  \label{tab:basis-C2-ext}
  \begin{tabular}{*{13}{>{$}c<{$}}}
    \ct_1 & \ct_2 & \ct_3 & \ct_4 & \ct_5 & \ct_6 & \ct_7 & \ct_8 & \ct_9 & \ct_{10} & \ct_{11} & \ct_{12} & \ct_{13}\\\hline
    \beta\pi Z & \beta\pi H  & \beta\pi R & \beta\pi B & \beta\pi P & \tfrac12\beta\beta R & \tfrac12\beta\beta B & \tfrac12\beta\beta P & \tfrac12\pi\pi R & \tfrac12\pi\pi B & \tfrac12\pi\pi P & \eta \pi R & \eta  \pi B\\[10pt]
    \ct_{14} & \ct_{15} & \ct_{16} & \ct_{17} & \ct_{18} & \ct_{19} & \ct_{20} & \ct_{21} & \ct_{22} & \ct_{23} & \ct_{24} & \ct_{25} & \\\hline
    \eta \pi P & \eta \beta R & \eta \beta B & \eta \beta P & \zeta\pi R & \zeta\pi B & \zeta\pi P  & \zeta\beta R & \zeta\beta B & \zeta\beta P & \eta \zeta Z & \eta  \zeta H  & 
  \end{tabular}
\end{table}

\begin{table}[h!]
  \centering
  \caption{Basis for $C^3(\th;\tg)^{\s}$}
  \label{tab:basis-C3-ext}
  \begin{tabular}{*{14}{>{$}c<{$}}}
    \bt_1 & \bt_2 & \bt_3 & \bt_4 & \bt_5 & \bt_6 & \bt_7 & \bt_8 & \bt_9 & \bt_{10} & \bt_{11}  \\\hline
    \eta \zeta\beta R & \eta \zeta\beta B & \eta \zeta\beta P & \eta \zeta\pi R & \eta \zeta\pi B & \eta \zeta\pi P  & \eta \beta\pi H  & \eta \beta\pi Z & \zeta\beta\pi H  & \zeta\beta\pi Z & \beta\beta\beta H  \\[10pt]
    \bt_{12}  & \bt_{13}  & \bt_{14} &  \bt_{15} & \bt_{16} & \bt_{17} & \bt_{18} & \bt_{19} & \bt_{20} & \bt_{21} & \bt_{22} \\\hline
\beta\beta\beta Z & \beta\beta\pi H  & \beta\beta\pi Z & \beta\pi\pi H  & \beta\pi\pi Z & \pi\pi\pi H  & \pi\pi\pi Z & \eta \beta\beta R & \eta \beta\beta B & \eta \beta\beta P & \eta \beta\pi R \\[10pt]
    \bt_{23} & \bt_{24} & \bt_{25}  & \bt_{26}  & \bt_{27}  & \bt_{28} & \bt_{29} & \bt_{30} & \bt_{31} & \bt_{32} & \bt_{33} \\\hline
    \eta \beta\pi B & \eta \beta\pi P & \eta \pi\pi R & \eta \pi\pi B & \eta \pi\pi P &  \zeta\beta\beta R & \zeta\beta\beta B & \zeta\beta\beta P & \zeta\beta\pi R & \zeta\beta\pi B & \zeta\beta\pi P\\[10pt]    
    \bt_{34} & \bt_{35} & \bt_{36} & \bt_{37} & \bt_{38} & \bt_{39}  & \bt_{40}  & \bt_{41}  & \bt_{42} & & \\\hline
    \zeta\pi\pi R & \zeta\pi\pi B & \zeta\pi\pi P & \beta\pi\beta R & \beta\pi\beta B & \beta\pi\beta P & \beta\pi\pi R & \beta\pi\pi B & \beta\pi\pi P & &                                    
  \end{tabular}
\end{table}

The Chevalley--Eilenberg differential is defined on generators by
\begin{equation}
  \d\zeta = - \beta^i \pi^i \qquad \d B_i = -\pi^i Z \qquad \d P_i = \beta^i Z \qquad\text{and}\qquad \d R_i = -\epsilon_{ijk} (\beta^j B_k + \pi^j P_k),
\end{equation}
and zero elsewhere. From these we can calculate the differential on cochains. Using the notation in the above tables of cochains, the non-zero differentials are:
\begin{equation}
  \label{eq:CE-static-ce-explicit}
  \begin{aligned}[m]
    \d \at_1 &= -\ct_2 \\
    \d \at_2 &= -\ct_1 \\
    \d \at_5 &= \ct_5 + 2 \ct_7\\
    \d \at_6 &= \ct_1 \\
    \d \at_8 &= \ct_4 + 2 \ct_{11}\\
    \d \at_{10} &= \ct_1\\
  \end{aligned}
  \qquad
  \begin{aligned}[m]
    \d \ct_3 &= -\bt_{38} - \bt_{42}\\
    \d \ct_4 &= -\bt_{16}\\
    \d \ct_5 &= \bt_{14}\\
    \d \ct_6 &= \bt_{39}\\
    \d \ct_7 &= -\tfrac12 \bt_{14}\\
    \d \ct_8 &= \tfrac12 \bt_{12}\\
    \d \ct_9 &= -\bt_{41}
  \end{aligned}
  \qquad
  \begin{aligned}[m]
    \d \ct_{10} &= -\tfrac12 \bt_{18}\\
    \d \ct_{11} &= \tfrac12 \bt_{16}\\
    \d \ct_{12} &= -\bt_{23} - \bt_{27}\\
    \d \ct_{14} &= -\bt_8\\
    \d \ct_{15} &= -\bt_{20} - \bt_{24}\\
    \d \ct_{16} &= -\bt_8\\
    \d \ct_{18} &= - \bt_{32} - \bt_{36} - \bt_{40}
  \end{aligned}
  \qquad
  \begin{aligned}[m]
    \d \ct_{19} &= -\bt_{41}\\
    \d \ct_{20} &= -\bt_{10} - \bt_{42}\\
    \d \ct_{21} &= -\bt_{29} - \bt_{33} - \bt_{37}\\
    \d \ct_{22} &= -\bt_{10} - \bt_{38}\\
    \d \ct_{23} &= -\bt_{39}\\
    \d \ct_{24} &= \bt_8\\
    \d \ct_{25} &= \bt_7.
  \end{aligned}
\end{equation}

Finally, we work out (the restriction of) the Nijenhuis--Richardson bracket
\begin{equation}
  [\![-,-]\!] : C^2(\h;\g)^{\s} \times C^2(\h;\g)^{\s} \to C^3(\h;\g)^{\s}
\end{equation}
on the above basis of cochains. Although not all of the brackets
appear in our calculations, we list the non-zero ones here for
completeness and in order to allow others to reproduce our
calculations.
\begin{equation}\tiny
  \label{eq:NR-ext-static}
  \begin{aligned}[m]
    [\![\ct_1, \ct_4]\!] &= \bt_{16}\\
    [\![\ct_1, \ct_5]\!] &= -\bt_{14}\\
    [\![\ct_1, \ct_7]\!] &= \tfrac12 \bt_{14}\\
    [\![\ct_1, \ct_8]\!] &= -\tfrac12 \bt_{12}\\
    [\![\ct_1, \ct_{10}]\!] &= \tfrac12 \bt_{18}\\
    [\![\ct_1, \ct_{11}]\!] &= -\tfrac12 \bt_{16}\\
    [\![\ct_1, \ct_{14}]\!] &= \bt_8\\
    [\![\ct_1, \ct_{16}]\!] &= \bt_8\\
    [\![\ct_1, \ct_{18}]\!] &= \bt_{40}\\
    [\![\ct_1, \ct_{19}]\!] &= \bt_{41}\\
    [\![\ct_1, \ct_{20}]\!] &= \bt_{10} + \bt_{42}\\
    [\![\ct_1, \ct_{21}]\!] &= \bt_{37}\\
    [\![\ct_1, \ct_{22}]\!] &= \bt_{10} + \bt_{38}\\
    [\![\ct_1, \ct_{23}]\!] &= \bt_{39}\\
    [\![\ct_1, \ct_{24}]\!] &= -\bt_8\\
    [\![\ct_1, \ct_{25}]\!] &= -\bt_7\\
    [\![\ct_2, \ct_4]\!] &= \bt_{15}\\
    [\![\ct_2, \ct_5]\!] &= -\bt_{13}\\
    [\![\ct_2, \ct_7]\!] &= \tfrac12 \bt_{15}\\
    [\![\ct_2, \ct_8]\!] &= -\tfrac12 \bt_{11}\\
    [\![\ct_2, \ct_{10}]\!] &= \tfrac12 \bt_{17}\\
    [\![\ct_2, \ct_{11}]\!] &= -\tfrac12 \bt_{15}\\
    [\![\ct_2, \ct_{12}]\!] &= \bt_{40}\\
    [\![\ct_2, \ct_{13}]\!] &= \bt_{41}\\
    [\![\ct_2, \ct_{14}]\!] &= \bt_7 + \bt_{42}\\
    [\![\ct_2, \ct_{15}]\!] &= \bt_{37}\\
    [\![\ct_2, \ct_{16}]\!] &= \bt_7 + \bt_{38}\\
    [\![\ct_2, \ct_{17}]\!] &= \bt_{39}\\
    [\![\ct_2, \ct_{20}]\!] &= \bt_9\\
    [\![\ct_2, \ct_{22}]\!] &= \bt_9
  \end{aligned}
  \qquad
    \begin{aligned}[m]
    [\![\ct_2, \ct_{24}]\!] &= \bt_{10}\\
    [\![\ct_2, \ct_{25}]\!] &= \bt_9\\
    [\![\ct_3, \ct_4]\!] &= -\bt_{40}\\
    [\![\ct_3, \ct_5]\!] &= \bt_{37}\\
    [\![\ct_3, \ct_7]\!] &= -\bt_{37}\\
    [\![\ct_3, \ct_{11}]\!] &= \bt_{37}\\
    [\![\ct_3, \ct_{13}]\!] &= \bt_{25}\\
    [\![\ct_3, \ct_{14}]\!] &= \bt_{22}\\
    [\![\ct_3, \ct_{16}]\!] &= \bt_{22}\\
    [\![\ct_3, \ct_{17}]\!] &= \bt_{19}\\
    [\![\ct_3, \ct_{19}]\!] &= \bt_{34}\\
    [\![\ct_3, \ct_{20}]\!] &= \bt_{31}\\
    [\![\ct_3, \ct_{22}]\!] &= \bt_{31}\\
    [\![\ct_3, \ct_{23}]\!] &= \bt_{28}\\
    [\![\ct_4, \ct_4]\!] &= -2 \bt_{41}\\
    [\![\ct_4, \ct_5]\!] &= \bt_{38}-\bt_{42}\\
    [\![\ct_4, \ct_6]\!] &= \bt_{37}\\
    [\![\ct_4, \ct_8]\!] &= \bt_{39}\\
    [\![\ct_4, \ct_{11}]\!] &= \bt_{38}\\
    [\![\ct_4, \ct_{13}]\!] &= \bt_{26}\\
    [\![\ct_4, \ct_{14}]\!] &= \bt_{23}\\
    [\![\ct_4, \ct_{15}]\!] &= -\bt_{22}\\
    [\![\ct_4, \ct_{17}]\!] &= \bt_{20} - \bt_{24}\\
    [\![\ct_4, \ct_{19}]\!] &= \bt_{35}\\
    [\![\ct_4, \ct_{20}]\!] &= \bt_{32}\\
    [\![\ct_4, \ct_{21}]\!] &= -\bt_{31}\\
    [\![\ct_4, \ct_{23}]\!] &= \bt_{29} - \bt_{33}\\
    [\![\ct_5, \ct_5]\!] &= 2 \bt_{39}\\
    [\![\ct_5, \ct_7]\!] &= -\bt_{39}\\
    [\![\ct_5, \ct_9]\!] &= -\bt_{40}
  \end{aligned}
  \qquad
  \begin{aligned}[m]
    [\![\ct_5, \ct_{10}]\!] &= -\bt_{41}\\
    [\![\ct_5, \ct_{11}]\!] &= \bt_{39} - \bt_{42}\\
    [\![\ct_5, \ct_{12}]\!] &= -\bt_{22}\\
    [\![\ct_5, \ct_{13}]\!] &= \bt_{27}-\bt_{23}\\
    [\![\ct_5, \ct_{16}]\!] &= \bt_{24}\\
    [\![\ct_5, \ct_{17}]\!] &= \bt_{21}\\
    [\![\ct_5, \ct_{18}]\!] &= -\bt_{31}\\
    [\![\ct_5, \ct_{19}]\!] &= \bt_{36}-\bt_{32}\\
    [\![\ct_5, \ct_{22}]\!] &= \bt_{33}\\
    [\![\ct_5, \ct_{23}]\!] &= \bt_{30}\\
    [\![\ct_6, \ct_{10}]\!] &= \bt_{40}\\
    [\![\ct_6, \ct_{13}]\!] &= \bt_{22}\\
    [\![\ct_6, \ct_{16}]\!] &= \bt_{19}\\
    [\![\ct_6, \ct_{19}]\!] &= \bt_{31}\\
    [\![\ct_6, \ct_{22}]\!] &= \bt_{28}\\
    [\![\ct_7, \ct_{10}]\!] &= \bt_{41}\\
    [\![\ct_7, \ct_{13}]\!] &= \bt_{23}\\
    [\![\ct_7, \ct_{15}]\!] &= \tfrac12 \bt_{19}\\
    [\![\ct_7, \ct_{16}]\!] &= -\tfrac12 \bt_{20}\\
    [\![\ct_7, \ct_{17}]\!] &= -\tfrac12 \bt_{21}\\
    [\![\ct_7, \ct_{19}]\!] &= \bt_{32}\\
    [\![\ct_7, \ct_{21}]\!] &= -\tfrac12 \bt_{28}\\
    [\![\ct_7, \ct_{22}]\!] &= \tfrac12 \bt_{29}\\
    [\![\ct_7, \ct_{23}]\!] &= -\tfrac12 \bt_{30}\\
    [\![\ct_8, \ct_9]\!] &= -\bt_{37}\\
    [\![\ct_8, \ct_{10}]\!] &= \bt_{42} - \bt_{38}\\
    [\![\ct_8, \ct_{11}]\!] &= -\bt_{39}\\
    [\![\ct_8, \ct_{12}]\!] &= -\tfrac12 \bt_{19}\\
    [\![\ct_8, \ct_{13}]\!] &= \bt_{24} - \tfrac12 \bt_{20}
  \end{aligned}
  \qquad
  \begin{aligned}[m]
    [\![\ct_8, \ct_{14}]\!] &= -\tfrac12 \bt_{21}\\
    [\![\ct_8, \ct_{16}]\!] &= \bt_{21}\\
    [\![\ct_8, \ct_{18}]\!] &= -\tfrac12 \bt_{28}\\
    [\![\ct_8, \ct_{19}]\!] &= \bt_{33} - \tfrac12 \bt_{29}\\
    [\![\ct_8, \ct_{20}]\!] &= -\tfrac12 \bt_{30}\\
    [\![\ct_8, \ct_{22}]\!] &= \bt_{30}\\
    [\![\ct_9, \ct_{14}]\!] &= \bt_{25}\\
    [\![\ct_9, \ct_{17}]\!] &= \bt_{22}\\
    [\![\ct_9, \ct_{20}]\!] &= \bt_{34}\\
    [\![\ct_9, \ct_{23}]\!] &= \bt_{31}\\
    [\![\ct_{10}, \ct_{14}]\!] &= \bt_{26}\\
    [\![\ct_{10}, \ct_{15}]\!] &= -\tfrac12 \bt_{25}\\
    [\![\ct_{10}, \ct_{16}]\!] &= -\tfrac12 \bt_{26}\\
    [\![\ct_{10}, \ct_{17}]\!] &= \bt_{23} - \tfrac12 \bt_{27}\\
    [\![\ct_{10}, \ct_{20}]\!] &= \bt_{35}\\
    [\![\ct_{10}, \ct_{21}]\!] &= -\tfrac12 \bt_{34}\\
    [\![\ct_{10}, \ct_{22}]\!] &= -\tfrac12 \bt_{35}\\
    [\![\ct_{10}, \ct_{23}]\!] &= \bt_{32} - \tfrac12 \bt_{36}\\
    [\![\ct_{11}, \ct_{12}]\!] &= -\tfrac12 \bt_{25}\\
    [\![\ct_{11}, \ct_{13}]\!] &= -\tfrac12 \bt_{26}\\
    [\![\ct_{11}, \ct_{14}]\!] &= \tfrac12 \bt_{27} \\
    [\![\ct_{11}, \ct_{17}]\!] &= \bt_{24}\\
    [\![\ct_{11}, \ct_{18}]\!] &= -\tfrac12 \bt_{34} \\
    [\![\ct_{11}, \ct_{19}]\!] &= -\tfrac12 \bt_{35} \\
    [\![\ct_{11}, \ct_{20}]\!] &= \tfrac12 \bt_{36} \\
    [\![\ct_{11}, \ct_{23}]\!] &= \bt_{33}\\
    [\![\ct_{12}, \ct_{20}]\!] &= -\bt_4
  \end{aligned}
  \qquad
    \begin{aligned}[m]
    [\![\ct_{12}, \ct_{23}]\!] &= -\bt_1\\
    [\![\ct_{12}, \ct_{25}]\!] &= \bt_4\\
    [\![\ct_{13}, \ct_{20}]\!] &= -\bt_5\\
    [\![\ct_{13}, \ct_{21}]\!] &= \bt_4\\
    [\![\ct_{13}, \ct_{22}]\!] &= \bt_5\\
    [\![\ct_{13}, \ct_{23}]\!] &= \bt_6 - \bt_2\\
    [\![\ct_{13}, \ct_{25}]\!] &= \bt_5\\
    [\![\ct_{14}, \ct_{18}]\!] &= \bt_4\\
    [\![\ct_{14}, \ct_{19}]\!] &= \bt_5\\
    [\![\ct_{14}, \ct_{23}]\!] &= -\bt_3\\
    [\![\ct_{14}, \ct_{25}]\!] &= \bt_6\\
    [\![\ct_{15}, \ct_{19}]\!] &= -\bt_4\\
    [\![\ct_{15}, \ct_{22}]\!] &= -\bt_1\\
    [\![\ct_{15}, \ct_{25}]\!] &= \bt_1\\
    [\![\ct_{16}, \ct_{19}]\!] &= -\bt_5\\
    [\![\ct_{16}, \ct_{21}]\!] &= \bt_1\\
    [\![\ct_{16}, \ct_{23}]\!] &= \bt_3\\
    [\![\ct_{16}, \ct_{25}]\!] &= \bt_2\\
    [\![\ct_{17}, \ct_{18}]\!] &= \bt_1\\
    [\![\ct_{17}, \ct_{19}]\!] &= \bt_2 - \bt_6\\
    [\![\ct_{17}, \ct_{20}]\!] &= \bt_3\\
    [\![\ct_{17}, \ct_{22}]\!] &= -\bt_3\\
    [\![\ct_{17}, \ct_{25}]\!] &= \bt_3\\
    [\![\ct_{18}, \ct_{24}]\!] &= \bt_4\\
    [\![\ct_{19}, \ct_{24}]\!] &= \bt_5\\
    [\![\ct_{20}, \ct_{24}]\!] &= \bt_6\\
    [\![\ct_{21}, \ct_{24}]\!] &= \bt_1\\
    [\![\ct_{22}, \ct_{24}]\!] &= \bt_2\\
    [\![\ct_{23}, \ct_{24}]\!] &= \bt_3
  \end{aligned}
\end{equation}


\providecommand{\href}[2]{#2}\begingroup\raggedright\endgroup

\end{document}